\documentclass[doublespacing]{elsart}
\usepackage{icarus}
\usepackage{natbib}

\bibpunct{(}{)}{;}{a}{,}{,}

\usepackage{epsfig}
\usepackage{graphics,color}
\begin{document} 
\begin{frontmatter}



\title{The Role of Ejecta\\
in the Small Crater Populations\\
on the Mid-Sized Saturnian Satellites}


\author[bierhaus]{Edward B.\ Bierhaus},
\author[dones]{Luke Dones},
\author[alvarellos]{Jos\'e Luis Alvarellos}, and
\author[zahnle]{Kevin Zahnle}

\address[bierhaus]{Lockheed Martin Space Systems Company,
                                Denver, CO 80201 (U.S.A.)}
\address[dones]{Southwest Research Institute,
                                Boulder, CO 80302 (U.S.A.)}
\address[alvarellos]{Loral Space Systems, 
                                Palo Alto, CA 94303 (U.S.A.)}
\address[zahnle]{NASA Ames, 
                                Moffett Field, CA 94035 (U.S.A.)}

\begin{center}
\scriptsize
Copyright \copyright\ 2011 Edward B.\ Bierhaus, Luke Dones,
Jos\'e Alvarellos, and Kevin Zahnle
\end{center}


%
%
%
%
%


\end{frontmatter}



\begin{flushleft}
\vspace{1cm}
Number of pages: \pageref{lastpage} \\
Number of tables: \ref{lasttable}\\
Number of figures: \ref{lastfig}\\
\end{flushleft}


\begin{pagetwo}{Ejecta and small craters in the Saturnian system}

Edward B.\ Bierhaus\\
Lockheed Martin Space Systems Company\\
MS S8110\\
PO Box 179\\
Denver, CO 80201\\

\end{pagetwo}

\begin{abstract}
We find evidence, by both observation and analysis, that crater
ejecta play an important role in the small crater (less than a
few km) populations on the Saturnian satellites, and more broadly,
on cratered surfaces throughout the Solar System.  We measure crater
populations in Cassini images of Enceladus, Rhea, and Mimas, 
focusing on image data with scales less than 500~m/pixel.  We use
recent updates to crater scaling laws and their constants \citep{housen2011}
to estimate the amount of mass
ejected in three different velocity ranges: (i) greater than escape
velocity, (ii) less than escape velocity and faster than the minimum
velocity required to make a secondary crater ($v_{\rm min}$), and (iii),
velocities less than $v_{\rm min}$.  Although the vast majority of mass on
each satellite is ejected at speeds less than $v_{\rm min}$, our
calculations demonstrate that the differences in mass available in the
other two categories should lead to observable differences in the
small crater populations; the predictions are borne out by the
measurements we have made to date.  In particular, Rhea, Tethys,
and Dione have sufficient surface gravities to retain ejecta 
moving fast enough to make secondary crater populations.  The
smaller satellites, such as Enceladus but especially Mimas, are 
expected to have little or no
traditional secondary populations because their escape velocities
are near the threshold velocity necessary to make a secondary crater.
Our work clarifies why the Galilean satellites have extensive
secondary crater populations relative to the Saturnian satellites.
The presence, extent, and sizes of sesquinary craters (craters formed 
by ejecta that
escape into temporary orbits around Saturn before re-impacting
the surface \citep{dobro2004, alvarellos2005, zahnle2008}
is not yet well understood.  Finally, our work provides further
evidence for a ``shallow" size-frequency distribution (slope
index of $\sim 2$ for a differential power-law) for comets
a few kilometers diameter and smaller.
\end{abstract}


\section{Introduction}
Without access to planetary samples from outer Solar System objects,
cataloguing the size-frequency distribution (SFD) and spatial
distribution of impact craters is a common means to determine surface
ages.  Constraining age on an object, and between objects, is
critical to establish a coherent and consistent chronology of events
within a planetary system, or across the Solar System.
Combining the measured crater SFD with known or modeled crater
formation rates provides a means to estimate absolute age.

\begin{figure}
\begin{center}
\epsfxsize=100mm
\includegraphics[width=5truein]{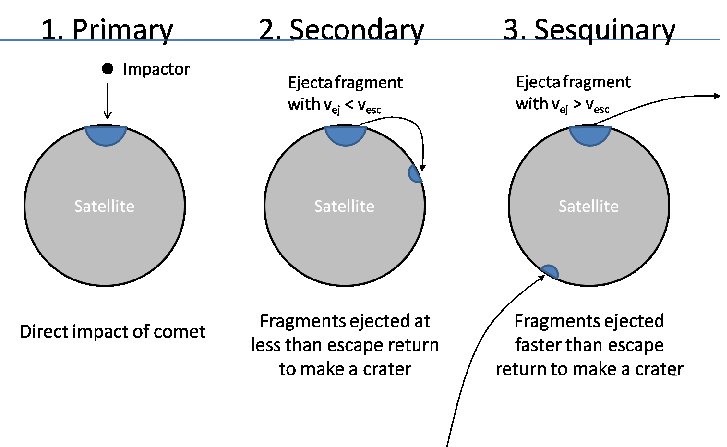}
\caption[Schematic of crater types.]
{A simple schematic of the three types of impact craters considered
in this paper.  (1) Primary craters are caused by the direct impact of
a comet or an asteroid.  (2) Secondary craters are caused by ejecta
from a primary crater that travels ballistically to some distance
away, and which impacts the surface with sufficient speed to form
a crater.  (3) Sesquinary craters are caused by ejecta that initially
escapes the satellite, and go into temporary orbit around the
planet.  Some time later the material the material impacts a satellite (usually the moon from which the material was ejected) to form a crater.}
\label{fig:types}
\end{center}
\end{figure}

To derive accurate ages using impact craters, in either the relative
or absolute sense, one must first determine the sources of impactors
that make craters.  Impact craters can be primary, secondary or
sesquinary; see Figure~\ref{fig:types}.  Primary craters are made by 
direct impact of comets
or asteroids.  Secondary craters are the result of essentially ballistic
trajectories of ejecta from the primary crater to some distance away.
For typical impact speeds of heliocentric comets onto
Saturnian satellites of several to 30~km/s \citep{zahnle2003, dones2009},
ejecta can be launched at speeds from a few hundred m/s to several km/s.  
When an
ejectum is launched at a speed faster than the escape velocity of a
moon, it can go into orbit about the planet.  Most escaped ejecta
are eventually swept up by the source moon, 
but the orbits of some
escaped ejecta can be sufficiently perturbed, or the original
ejection velocity so high, that the ejecta will
impact another satellite \citep{alvarellos2005,alvarellos2002}. 
In either case, craters formed by
ejecta that initially escape their parent object are called
sesquinary (``$1 \frac{1}{2}$-ary";  formerly ``poltorary'') craters \citep{dobro2004, zahnle2008}.  
Because secondary and sesquinary craters
are products of primary craters, and because the larger (and
therefore generally older) primary craters create the most ejecta,
older terrains will have the greatest number of craters of all types.
This introduces uncertainty in the number of primary craters, which
is the only kind to be trusted chronometrically \citep{mcewen2006}.

Most Voyager-era studies of craters on the Saturnian moons identified two basic
classes, called Population I and Population II \citep{smith1981, smith1982}. Population II
follows a steeper size-frequency distribution than Population I
(i.e., in relative terms, Population II has a smaller number of
large craters). One hypothesis proposed (e.g., \cite{chapman1986}) 
that heliocentric
comets are responsible for Population I, while planetocentric debris from cratering or catastrophic disruption of moons is
responsible for Population II.  A second hypothesis (e.g., \citet{strom1982}) proposed
that planetocentric debris largely is responsible for both populations. A third \citep{plescia1985} suggested that Population II was created by late-arriving bodies from heliocentric orbit. The
existence of two populations was questioned by Lissauer
et al. (1988), who argued that the Voyager data on small craters might be consistent with (near-)saturation equilibrium produced by heliocentric impactors \citep{hartmann1984}. For our purposes, the key point is that ever since craters were first seen on the Saturnian moons, the question of whether the impactors were heliocentric, planetocentric, or both has been controversial.

Since the Voyager flybys, a number of developments have taken place 
(see reviews by \citet{chapman1986, schenk2004, dones2009}).  First is 
the confirmation of the \cite{shoemaker1981} and \citet{shoemaker1982}
prediction that heliocentric ``ecliptic'' comets should dominate primary 
impacts on planets and regular satellites in the outer Solar System 
\citep{levison1997, zahnle1998, zahnle2003}, and the discovery of their 
putative source region, the Kuiper Belt/Scattered Disk \citep{barucci2008}.  
Second is the
resurrection of Shoemaker's \citep{shoemaker1965} observation that secondary craters can be a major component of the small crater population \citep{bierhaus2005, mcewen2005, mcewen2006}.  Third is the detailed modeling of the dynamical behavior of escaped ejecta in multi-body systems (such as the Saturnian system), leading to the first quantitative estimate of the effect of sesquinary
ejecta launched from large craters on the mid-sized moons \citep{moore2004} on the cratering population \citep{alvarellos2005}.  Fourth is the ongoing laboratory and analytical work into the physics of crater ejecta \citep{melosh1984, housen2011}.  
Fifth is the possible origin of irregular satellites
of the giant planets, and possible consequences of their collisional evolution for the cratering record on inner satellites \citep{bottke2010}.  
The capstone, of course, is the presence of the Cassini mission at Saturn, and the imaging (ISS) data that enable a significant improvement of the catalogue of craters. Crater counts on the Saturnian moons using ISS images have been published for the following moons: Phoebe and Iapetus \citep{porco2005}; Hyperion and Phoebe \citep{thomas2007}; basins on the leading face of Iapetus \citep{giese2008}; Enceladus \citep{porco2006, kirchoff2009}; and Mimas, Tethys, Dione, Rhea, Iapetus, and Phoebe \citep{kirchoff2010}. Counts for Titan, using Cassini's radar instrument, have been published by \citet{lorenz2007, wood2010}, and \citet{neish2011}.

The first development cited above -- the prediction that ecliptic comets are the primary impactors on the regular satellites of the giant planets -- is based, in large part, on the sheer numbers of comets that seem to wander the outer Solar System \citep{zahnle1998, zahnle2003}.  Estimates of the impactor populations at each planet mostly rely on interpolation between observations of Jupiter-family comets (generally seen well interior to Jupiter's orbit, and prone to disintegrate for reasons that are not understood) and counts of (large) Centaurs and Kuiper Belt Objects, corrected for discovery biases with the help of dynamical models \citep{dones2009}. Detailed comparison of crater SFDs and the independently known population of impactors, such as has been carried out for the Moon by \citet{marchi2009}, is not possible for satellites in the outer Solar System.  

On synchronously rotating satellites, ecliptic comets are much more likely to strike the moons' leading hemispheres \citep{shoemaker1982, horedt1984a, zahnle2001}. The expected ``apex-antapex asymmetries'' do not occur to the extent expected, although craters on bright terrains on Ganymede \citep{zahnle2001, schenk2004} and rayed craters on Saturnian satellites \citep{schenk2011} do exhibit smaller asymmetries in the expected sense. 
(Triton is just weird; see \citet{schenk2007} if you must know the details.) 
The absent or muted apex/antapex effect might be due to crater saturation, 
true polar wander, or some form of nonsynchronous rotation.
 In addition, moons closer to their parent planets should be more heavily cratered because of gravitational focusing \citep{smith1981, smith1982, zahnle2003}. This effect is not observed either, perhaps as a result of saturation, at least for small craters, due to an early era of heavy bombardment \citep{chapman1986, dones2009, richardson2009}. 

Alternatively, one can abandon the hypothesis that the impactors 
have a heliocentric origin, and posit that the craters are 
produced by planetocentric (Saturn-orbiting) bodies 
\citep{horedt1984b, neukum1985}. The planetocentric model has 
the advantage that planetocentric cratering only weakly
favors a satellite's leading or trailing side, depending on whether the
debris fall from outside or inside the moon's orbit, respectively \citep{horedt1984a, alvarellos2005}.
However, planetocentric bodies on crossing orbits are generally short-lived \citep{burns1998, alvarellos2005}, so they must be resupplied if they are to contribute to latter-day cratering. Ironically, the most plausible source of resupply is impact by heliocentric comets, which might produce copious fragments that rain back onto the moons. This realization motivates our study of the role of ejecta from cometary impacts in producing the observed crater populations.

The outline of our paper is as follows. In Section 2, we apply the cratering ejecta model of \citet{housen2011} to derive the expected mass available to form sesquinary and secondary craters on Saturn's mid-sized moons and Jupiter's three icy Galilean satellites. In Section 3, we describe the image processing we performed on the publicly available Cassini images we used to measure crater populations. In Section 4,  we present the size-frequency distributions of craters on terrains on Enceladus, Mimas, and Rhea. In Section 5, we compare the observed and predicted populations of sesquinary craters on the Saturnian moons. Section 6 summarizes our conclusions.   

\section{Estimating the Role of Ejecta as Impactors}
The crater formation and evolution process has been long approximated
by a set of scaling laws, anchored to physical reality by
laboratory-scale experiments, nuclear explosions, observations of
impact craters across the Solar System, and increasing sophistication
of numerical modeling.  Although the details of any one impact are
not captured by the scaling laws, they provide a reasonable
description of trends.

\begin{figure}
\begin{center}
\epsfxsize=100mm
\includegraphics[width=5truein]{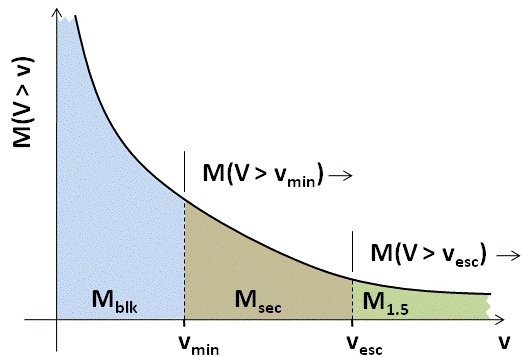}
\caption[Schematic of ejected mass.]
{A schematic of the amount of mass ejected faster than a
given velocity.  There is an inverse relationship between
mass and velocity, i.e.\ less mass is ejected at higher
velocities.  $v_{\rm min}$ is the minimum velocity required to
make a secondary,  while $v_{\rm esc}$ is the escape velocity for a
body.  Mass ejected faster than the escape velocity is the same
as the mass available to make sesquinary craters, or
$M(v>v^H_{\rm esc})=M_{1.5}$.  ($v^H_{\rm esc}$ is the modified
escape velocity, see text for discussion.)  Mass ejected faster 
than $v_{\rm min}$
but slower than $v_{\rm esc}$ is the mass available to make
secondary craters,  $M_{\rm sec}$ (unless $v_{\rm esc} < v_{\rm min}$, in which case no mass is available to make secondaries).  Mass ejected slower than
$v_{\rm min}$ is available to make an ejecta blanket,
$M_{\rm blk}$.}
\label{fig:ejectamass}
\end{center}
\end{figure}

For the current research, we focus on the mass-velocity relationship
of crater ejecta.  In particular, observations of laboratory-scale
cratering experiments \citep{cintala1999} reveal that there is an 
inverse relationship
between ejecta mass and ejection velocity, which is to say that less
mass is ejected at higher velocities, see Figure~\ref{fig:ejectamass}.  
Early stage ejecta from near the point of impact move the
fastest, but are the smallest portion of ejected mass.  Late-stage
ejecta are moving the slowest, but are the bulk of ejecta mass.
Drilling at terrestrial craters and observations of craters on
other surfaces indicate that the end of the crater excavation phase
is a transition to material that isn't so much ejected as it
is an overturned flap of the surface.

Using the most recent summary of laboratory measurements of cratering
ejecta from \citet{housen2011}, and impact velocities for ecliptic comets 
on the satellites from \citet{zahnle2003}, we estimate
the total and fractional ejecta masses for a 1~km diameter
cometary (density of 600~kg/$\mathrm{m}^3$) impactor on various Saturnian 
satellites.  We use a 1~km comet because
it is sufficiently large to generate ejecta masses that
contribute to the observable secondary and sesquinary crater populations,
but small enough that
such impacts should occur on geologically short timescales.
Table~\ref{tbl:satdata}
lists properties of the satellites used in our calculations.  To complete
the calculations, we must quantitatively determine the appropriate
velocity regimes that separate the ejecta blanket, secondaries,
and sesquinaries.  We discuss these velocities next.

\subsection{Making an Impact: $v_{\rm min}$}
\label{subsec:vmin}
We define $v_{\rm min}$ as the boundary between
the ejecta blanket and secondaries, i.e., it is the minimum velocity
at which an ejectum can make a secondary crater.  

We use observations of adjacent secondaries on Europa, and
the simple planar ballistics equation, to set
plausible values of $v_{\rm min}$ for icy surfaces.  The
ballistics equation is:

\begin{equation}
r = \frac{v^2 \sin 2\theta}{g}
\end{equation}
\noindent
where $r$ is range, $v$ is ejection velocity, $\theta$
is the ejection angle (measured relative to local horizontal),  
and $g$ is the surface gravity.
Measuring $r$ enables us to calculate $v$:
\begin{equation}
v = \left( \frac{g \, r}{\sin 2\theta} \right)^{1/2}
\label{eqn:vej}
\end{equation}

\begin{figure}
\begin{center}
\epsfxsize=100mm
\includegraphics[width=5truein]{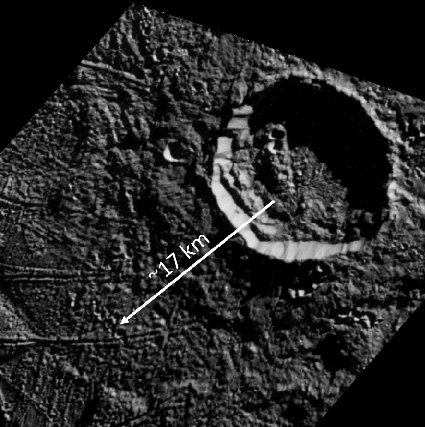}
\caption[Europan crater Rhiannon.]
{A Galileo image of Rhiannon, which is a
$\sim 15$~km diameter primary crater on Europa.  The nearest distinct
secondary craters are about 17~km from a point half-way between
Rhiannon's center and the crater rim.}
\label{fig:rhiannon}
\end{center}
\end{figure}

First, we
examine Rhiannon, a 15~km diameter crater on Europa; see
Figure~\ref{fig:rhiannon}.  Rhiannon is the smallest primary
on Europa imaged with sufficient resolution to see its adjacent
secondary crater field.  (Galileo imaged smaller primaries on
Europa, but not with sufficient resolution to determine whether
or not they have secondaries.)  The point from which the
fragments that make these secondaries originate is somewhat
uncertain.  The late ejecta come from the region close
to the crater rim.  The final crater diameter is larger than
the transient diameter due to collapse of the crater after
the excavation phase is complete.  We use a point roughly
halfway between the crater center and the final crater rim
as a reasonable location for the origin of the secondary
fragments.  The distance between our presumed origin point
and the first appearance of the adjacent secondary crater
population is about 17~km.  
Substituting $r=17$~km, $\theta = 45^{\circ}$,
and $g = 1.31$~m/$\mathrm{s^2}$ for Europa into
Equation~\ref{eqn:vej} gives $v \sim 150$~m/s.

\begin{figure}
\begin{center}
\epsfxsize=100mm
\includegraphics[width=5truein]{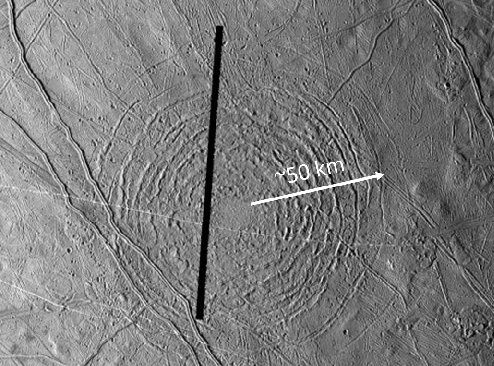}
\caption[Europan crater Tyre.]
{A Galileo image of Tyre, which is a
45 to 50~km diameter primary crater on Europa, depending on which
feature is the true crater rim.  The secondary crater population
begins about 50~km from the central portion of the crater.  The
black bar is a data gap.}
\label{fig:tyre}
\end{center}
\end{figure}

Next, we examine Tyre, the largest known primary
impact structure on Europa; see Figure~\ref{fig:tyre}.  Depending
on the location of the crater rim, Tyre is $45-50$~km diameter.
Following the procedure for measuring range we defined for
Rhiannon, we find that secondaries begin to appear about
50~km from the estimated point of origin.  Using 50~km in
Equation~\ref{eqn:vej} gives $v \sim 250$~m/s.

Thus, in our subsequent analysis, we examine two
different cases for the minimum velocity required to make 
secondaries, $v_{\rm min} = 150$~m/s and $v_{\rm min} = 250$~m/s.
Although the actual value could differ for the Saturnian
satellites, the general trends described here would not change.

\subsection{The Great Escape: $v_{\rm esc}^H$}

\citet{alvarellos2002,alvarellos2005} clarified that the relevant velocity
to escape a moon in orbit around a planet isn't the classical escape
velocity appropriate for an isolated body, but rather the velocity required 
to reach the moon's Hill Radius, which is:

\begin{equation}
R_H = a_m \left[ \frac{M_m}{3(M_m + M_p)} \right]^{1/3}
\label{eqn:rhill}
\end{equation}

\noindent
where $a_m$ is the moon's semi-major axis, $M_m$ is the mass
of the moon, and $M_p$ is the mass of the planet.  The velocity
necessary to reach $R_H$ is then:

\begin{equation}
v^H_{\rm esc} = v_{\rm esc} \left( \frac{R_H^2 - R_H R_m}{R_H^2 - R_m^2/2} \right)^{1/2},
\label{eqn:vhill}
\end{equation}

\noindent
where $R_m$ is the moon's radius, $v_{\rm esc} = \sqrt{2GM_m/R_m}$ 
is the classical escape velocity, and $M_m$ is the moon's mass. 
When estimating the amount of mass that escapes a moon, we use
$v_{\rm esc}^H$ rather than $v_{\rm esc}$.  In general,
$v_{\rm esc}^H < v_{\rm esc}$.

\subsection{Scaling Law Calculations}
\label{sec:calcs}

Our calculations are based on the expressions and definitions within
\citet{housen2011}.  First is the expression for the cumulative mass
ejected faster than a given velocity for a crater in the gravity
regime, defined in terms of impactor properties:

\begin{equation}
\frac{M(v>\hat{v})}{m_i} = C_{4} \left( \frac{\hat{v}}{v_i}\right)^{-3\mu} \left( \frac{\rho_t}{\rho_i} \right)^{1-3\nu}
\label{eqn:hh11sym}
\end{equation}
\noindent
where $C_4$, $\nu$, and $\mu$ are constants, and  $M(v>\hat{v})$ is the mass 
ejected faster than velocity $\hat{v}$, $v_i$ is the impactor velocity,
$\rho_t$ is the density of the target's crust, and $\rho_i$ is the impactor density.
\citet{housen2011} indicate that $\nu$ is about 0.4 regardless
of target type.  The value of the exponent $\mu$ is expected to 
lie between $\frac{1}{3}$(``momentum scaling'') and $\frac{2}{3}$ 
(``energy scaling'').  For the materials tabulated by \citet{housen2011}, 
$\mu$ ranges between 0.35 for perlite/sand mixture to 0.55 for water 
and rock.  The values of $\mu$ and $C_4$ for ice are not given, but 
K.\ Housen (personal communication, 2011) recommends using the values 
for weakly cemented basalt as the closest approximation
for ice ($\mu=0.46$, $C_{4}=6.72 \times 10^{-3}$). 
We set $\rho_t=900$~kg/$\mathrm{m}^3$, appropriate for nonporous ice. 


For a 1~km diameter comet with a density of 600~kg/$\mathrm{m}^3$,
the impactor's mass is $3.14 \times 10^{11}$~kg.  We use
\citet{zahnle2003} as the source for typical values of $v_i$ for ecliptic comets
striking the different satellites.  Substituting numerical values
into Equation~\ref{eqn:hh11sym}, we obtain the following for
ejecta masses on icy satellites:

\begin{equation}
M(v>\hat{v}) = 6.72 \times 10^{-3} m_i \left( \frac{\hat{v}}{v_i} \right)^{-1.38} \left( \frac{\rho_t}{\rho_i} \right)^{-0.2}
\label{eqn:hh11num}
\end{equation}

We will henceforth write $M(v>\hat{v})$ as $M(\hat{v})$.
First we calculate the ejected mass that escapes the moon, $M(v^H_{\rm esc})$.
In our discussion, we assume that this mass is equivalent to the 
amount of mass available to make sesquinary craters, $M_{1.5}$.
\begin{equation}
M_{1.5} = M(v^H_{\rm esc}) = 6.72 \times 10^{-3} m_i\ \left( \frac{v^H_{\rm esc}}{v_i} \right)^{-1.38} \left( \frac{\rho_t}{\rho_i} \right)^{-0.2}
\end{equation}
\citet{alvarellos2005} demonstrated that, for the mid-sized 
Saturnian satellites,
most but not all of the sesquinary fragments re-impact their source moon.
Their results were bracketed by 81.5\% re-impact (ejecta from Odysseus crater
on Tethys) on the low end, and by 99.6\% re-impact (ejecta from Herschel crater
on Mimas) on the high end.

Next we calculate the mass available to make secondary craters, 
$M_{sec}$, which is:
\begin{equation}
M_{sec} = M(v_{\rm min}) - M_{1.5}
\end{equation}

\noindent
where $M(v_{\rm min})$ is the mass ejected faster than the minimum speed
necessary to make secondary craters, $v_{\rm min}$.  This value is likely
variable, dependent on target properties, but observations of Europan
secondaries provide some constraints on the minimum possible value.
See our discussion in Section~\ref{subsec:vmin}, where we identify
two plausible values for $v_{min}$ of 150~m/s and 250~m/s.

\begin{equation}
M(v_{\rm min}) = 6.72 \times 10^{-3} m_i \left( \frac{v_{\rm min}}{v_i} \right)^{-1.38} \left( \frac{\rho_t}{\rho_i} \right)^{-0.2}
\end{equation}

\noindent
Later we find it useful to compare these masses with the total
ejecta mass for a crater, $M_{\rm tot}$.
As \citet{housen2011} discuss, 
the volume of the final transient crater -- at the end of
the crater excavation phase but before modification due to wall
collapse, etc.\ -- has a corresponding mass:
\begin{equation}
M_{\rm crater} = k_{\rm crater} \rho_t R^3
\end{equation}
\noindent
where $k_{crater}$ is a constant, and $R$ is the radius of the
transient crater at the end of the excavation stage.  This value does
not correspond to an actual physical mass, since some significant
fraction of any crater is formed by compaction and/or inelastic
compression rather than
excavation.  K.\ Housen (personal communication, 2011) recommends that 
$M_{\rm tot}/M_{crater}$ is
about 0.5 for craters on icy satellites.
Thus, we use the following expression for the total mass ejected:
\begin{equation}
M_{\rm tot} = 0.5\, M_{\rm crater} = 0.5\, k_{\rm crater}\, \rho_t \, R^3,
\end{equation}
where we assume $k_{\rm crater} = 0.6$.

To calculate $R$, we used the
expression from \citet{housen2011} for the diameter of the
transient crater in the gravity regime, given a certain impactor:
\begin{equation}
D = 2 H_1 \left( \frac{\rho_t}{m_i} \right)^{-1/3}  \left( \frac{\rho_t}{\rho_i} \right)^{\frac{2+\mu-6 \nu}{3(2+\mu)}} \left( \frac{g a}{v_i^2} \right)^{- \frac{\mu}{2+\mu}},
\end{equation}
\noindent
where $H_1$ is a constant. K.\ Housen (personal communication, 2011)
recommends using \citet{housen2011}'s value for sand of $H_1 = 0.59$
because of the similarity in porosity, although there may be 
differences due to friction angle.  Substituting $\mu = 0.46$, $\nu = 0.4$, 
we have
\begin{equation}
D = 1.08 \left(\frac{v_i^2}{g}\right)^{0.187} \left(\frac{\rho_i}{\rho_t} \right)^{0.325} d^{0.813},
\label{eqn:craterrad}
\end{equation}
as compared with
\begin{equation}
D = 1.1 \left(\frac{v_i^2}{g}\right)^{0.217} \left(\frac{\rho_i \cos i}{\rho_t} \right)^{0.333} d^{0.783},
\end{equation}
the expression used by \citet{zahnle2003,alvarellos2008,zahnle2008}, 
where $i$ is the angle of incidence (0 for a vertical impact). These 
authors assume $\mu = 0.55$, appropriate for impacts into wet sand, 
rock, and water \citep{schmidt1987,housen2011}.  We incorporate 
impact angle into our calculations (Equations~\ref{eqn:hh11num}
and \ref{eqn:craterrad}) by assuming an average primary
impact angle of $45^{\circ}$, and scaling the impact velocities
in \cite{zahnle2003} by $\cos 45^{\circ} = 0.707$.

With $M_{\rm tot}$, it is a simple
exercise to calculate the mass available to make an ejecta blanket 
($M_{\rm blk}$):
\begin{equation}
M_{\rm blk} = M_{\rm tot} - M(v_{\rm min})
\end{equation}

Figure~\ref{fig:ejectamass} is a schematic illustration of these
relationships.  Any mass moving faster than $v^H_{\rm esc}$ is available
to make sesquinary craters.  Mass ejected faster than $v_{\rm min}$
but slower than $v^H_{\rm esc}$ makes secondary craters ($M_{sec}$).  Mass ejected
slower than $v_{\rm min}$ is available to make the ejecta blanket 
($M_{\rm blk}$).  The
smaller the difference between $v_{\rm min}$ and $v^H_{\rm esc}$, the less
mass is available to make secondary craters.  (For very small moons, we have $v_{\rm min} < v^H_{\rm esc}$, in which case {\it no} mass is available to make secondaries.) Conversely, the
greater the difference between the two values, the more mass is
available to make secondary craters.

The effective escape velocity $v^H_{\rm esc}$ is a function of the mass and radius (or bulk density and radius) of 
the target object, as well as the moon's distance from its planet (see Eqs.~\ref{eqn:rhill} and \ref{eqn:vhill}).
The minimum velocity of ejecta that can make secondaries, $v_{\rm min}$, is a function of the surface
strength.  Presumably $v_{\rm min}$ is greater for rocky objects than
for icy objects, although the difference may not be significant.
For the mid-sized icy satellites of Saturn, $v_{\rm esc}$ is
smaller than it is for the terrestrial planets (or the Galilean
satellites).

For purposes of comparison, 
we calculate $M_{\rm tot}$, $M_{\rm blk}$, $M_{sec}$, and $M_{1.5}$ for
a 1~km diameter comet impact on several mid-sized icy Saturnian satellites
and the three icy Galilean satellites.  
We use the impact velocities from \citet{zahnle2003}, and an
impactor density of 600~kg/$\mathrm{m}^3$.  

In all cases, the vast majority of the ejected mass resides in 
the ``ejecta blanket".  Table~\ref{tbl:150dat} lists the
absolute mass and fractional mass for each moon when $v_{\rm min}=150$~m/s,
the case which minimizes the amount of mass available for $M_{\rm blk}$.
Even in this case, between 95\% and about 99\% of the mass is in
$M_{\rm blk}$.  However, enough ejecta are launched at sufficient speeds
that such an impact should add to the observable secondary and
sesquinary crater populations.  To illustrate this point further,
Table~\ref{tbl:impmass} lists $M_{sec}$ and $M_{1.5}$ in terms of
the impactor mass.  Depending on the value of $v^H_{\rm esc}$ for each
moon, the primary impact can eject multiple impactor masses to
create secondaries and/or sesquinaries.

\begin{figure}
\begin{center}
\epsfxsize=100mm
\includegraphics[width=5truein]{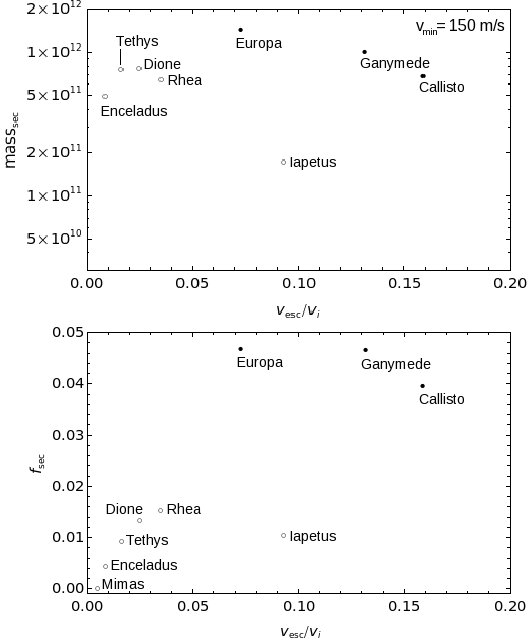}
\caption[Ejected mass to make secondaries for $v_{\rm min} = 150$~m/s.]
{A plot of $M_{\rm sec}$ (top panel) and $f_{\rm sec}$ (bottom panel), the
absolute and fractional ejected mass available to make
secondary craters for a 1 km cometary impactor, for
$v_{\rm min} = 150$~m/s.  The Saturnian satellites are
open circles, and the Galilean satellites are closed circles.
The mass available for secondaries on Mimas is zero for this
value of $v_{\rm min}$,
i.e.\ none of the ejecta retained by the satellite are moving
fast enough to make secondary craters.  The Galilean satellites,
and Europa in particular, have much more mass available for
secondary craters.}
\label{fig:secmassv150}
\end{center}
\end{figure}

\begin{figure}
\begin{center}
\epsfxsize=100mm
\includegraphics[width=5truein]{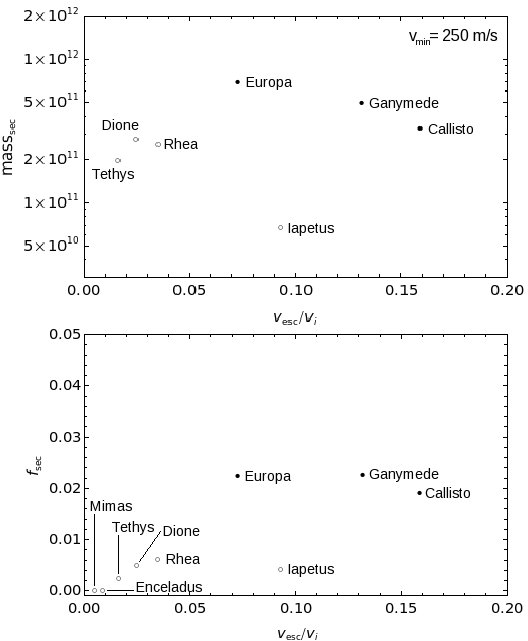}
\caption[Ejected mass to make secondaries for $v_{\rm min} = 250$~m/s.]
{A plot of $M_{\rm sec}$ (top panel) and $f_{\rm sec}$ (bottom panel), the 
absolute and fractional ejected mass available to make
secondary craters for a 1 km cometary impactor, for
$v_{\rm min} = 250$~m/s.  The Saturnian satellites are
open circles, and the Galilean satellites are closed circles.
The mass available for secondaries on both Mimas and Enceladus is zero 
for this value of $v_{\rm min}$,
i.e.\ none of the ejecta retained by the satellites are moving
fast enough to make secondary craters.}
\label{fig:secmassv250}
\end{center}
\end{figure}

Figures~\ref{fig:secmassv150} and \ref{fig:secmassv250} plot the 
fractional and absolute mass available to make secondary craters
on each of the satellites.  Mimas has no mass available to make
secondary craters in either case, because $v^H_{\rm esc} < v_{\rm min}$
for that moon, while the same is true for Enceladus in the
$v_{\rm min}=250$~m/s case.  In both cases, Europa has the highest
mass available for secondary craters.  Tethys, Dione, and Rhea
are the most similar to one another among the satellites addressed
in this study.  

\begin{figure}
\begin{center}
\epsfxsize=100mm
\includegraphics[width=5truein]{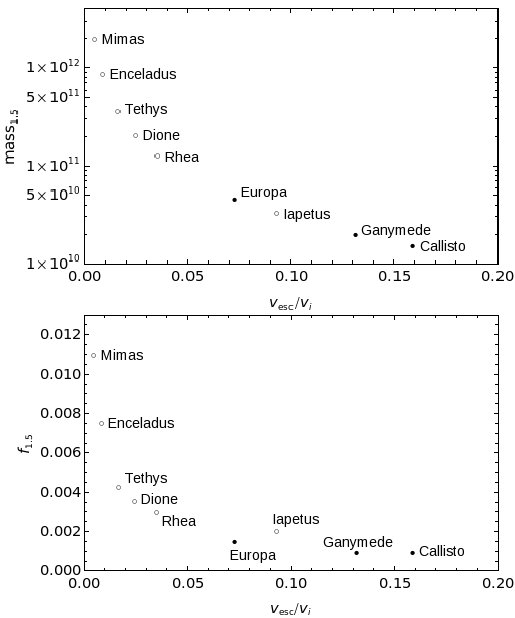}
\caption[Ejected mass to make sesquinaries.]
{A plot of $M_{1.5}$ (top panel) and $f_{1.5}$ (bottom panel), the 
absolute and fractional ejected mass available to make
sesquinary craters, for a 1 km cometary impactor. The Saturnian satellites
are represented by open circles, while the Galilean satellites are shown 
as closed circles.  $M_{1.5}$ is independent of $v_{\rm min}$.  Mimas and 
Enceladus have much more mass available to make sesquinary craters than
the other Saturnian satellites or the Galilean satellites.}
\label{fig:escmass}
\end{center}
\end{figure}

Figure~\ref{fig:escmass} plots the fractional and absolute mass
available to make sesquinary craters on each of the satellites.
$M_{1.5}$ is independent of $v_{\rm min}$, and so is the same
between the two values of $v_{\rm min}$ used here.  Mimas and
Enceladus have orders of magnitude more mass available in this
ejecta category than the Galilean satellites, and have significantly
greater amounts of sesquinary ejecta than the other mid-sized
Saturnian satellites.  

\subsection{Secondaries: Near and Far}
\label{subsec:nearfar}
Frequently secondaries are grouped into two basic populations:
adjacent and distant \citep{mcewen2006}.  Adjacent
secondaries are those that appear immediately outside the
continuous ejecta blanket of a primary crater, and are a
distinct high-spatial density annulus of small craters around
the primary.  The ``distant" secondary
is a broad classification, covering those secondaries that appear
several crater radii away, to those that are hundreds (or even
thousands) of km away \citep{mcewen2005,mcewen2006,preblich2007,
dundas2007,robbins2011}. Of course, an ejecta fragment with a given velocity will result
in different ballistic ranges on objects with different
surface gravities and sizes.  Formally, the
ballistic range $R$ of an ejecta fragment, launched from a point
on a sphere and traveling less than the escape speed of the
object, is: 

\begin{equation}
R = 2 \phi R_m
\end{equation}

\noindent
where $R_m$ is the radius of the moon, and $\phi$ is the half-angle
distance of travel, defined by:

\begin{equation}
\tan \phi = \frac{v_{\rm ej}^2 \sin \theta \cos \theta}{g R_m - v_{\rm ej}^2 \cos^2 \theta}
\end{equation}

\noindent
\citep{vickery1986}, where $v_{\rm ej}$ is the fragment ejection velocity, $\theta$ is the ejection 
angle, and $g$ is the surface gravity of the moon.

\begin{figure}
\begin{center}
\epsfxsize=100mm
\includegraphics[width=5truein]{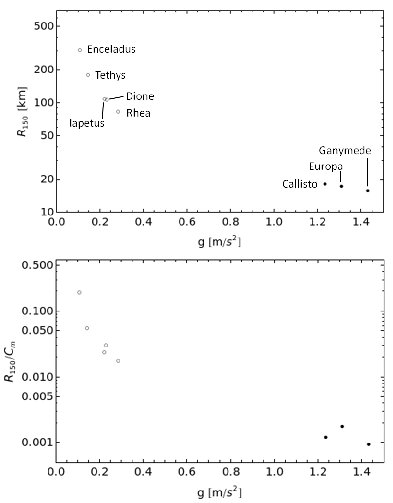}
\caption[Range for 150 m/s ejecta.]
{The top plot is the non-planar ballistic range in km on each of the 
moons for ejecta launched
at 150~m/s and a $45^{\circ}$ angle, as a function of the surface 
gravities of the satellites.  The bottom plot is that range, normalized 
to the circumference of each
satellite.  Mimas does not appear because ejecta launched at 150 m/s
escapes that moon.  The horizontal axes are the same in the two plots.
See text for further discussion.}
\label{fig:range150ms}
\end{center}
\end{figure}

\begin{figure}
\begin{center}
\epsfxsize=100mm
\includegraphics[width=5truein]{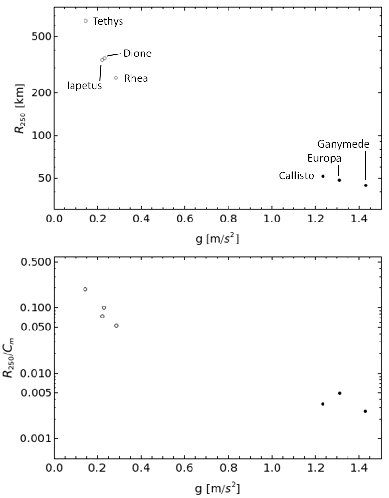}
\caption[Range for 250 m/s ejecta.]
{The same as Figure~\ref{fig:range150ms}, but for ejecta launched
at 250~m/s.  The bottom
plot is that range, normalized to the circumference of each
satellite.  Neither Mimas nor Enceladus appear because ejecta launched
at 250 m/s escapes these moons.  The horizontal axes are the same in 
the two plots.  See text for further discussion. }
\label{fig:range250ms}
\end{center}
\end{figure}

An illustrative comparison is to normalize the range by the circumference
of the body, i.e.\ $R/C_m$, where $C_m = 2 \pi R_m$ is the circumference of
a moon.  In Figure~\ref{fig:range150ms}, we plot $R$ and $R/C_m$ 
for $v_{\rm ej} = 150$~m/s,
i.e., the lower limit to $v_{\rm min}$, and in Figure~\ref{fig:range250ms} we plot the
same information for $v_{\rm ej} = 250$~m/s, the maximum value of $v_{\rm min}$
we use for the current analysis.

For $v_{\rm min}=150$~m/s, a secondary crater on Enceladus will form no
closer than over 300~km distant, and even for the more
massive Rhea, the closest secondary forms over 80~km away.
(In this model, no traditional secondaries form on Mimas for  $v_{\rm min}=150$~m/s.) 
This minimum range is a few percent of Rhea's circumference and almost 20\% of
Enceladus's circumference.

For $v_{\rm min}=250$~m/s, the closest secondaries move further away
from their parent primary.  (For $v_{\rm min}=250$~m/s,
the available mass to make secondaries is zero for both Mimas
and Enceladus.)  The closest secondaries on Rhea are over
200~km distant, while those on Dione and Iapetus are over 300~km
distant, and those on Tethys are over 600~km distant --
20\% the circumference of the moon.

On objects with small surface gravities, then,
there may not be ``adjacent" secondary populations in
the traditional sense.  Certainly we don't expect to
see dense, overlapping fields of secondary craters
around a large primary.  This is true for the
Saturnian satellites under discussion here, as well
as asteroids, Kuiper Belt Objects, and other bodies 
where $v_{\rm min}$ is close to, but still less than,
$v_{\rm esc}$ (or $v^H_{\rm esc}$).

\subsection{Discussion}
Based on examination of the tables and figures, we divide the satellites
into different groups based on the relative values of $M_{sec}$ and
$M_{1.5}$.  (Table~\ref{tbl:impmass} divides the satellites into groups with 
horizontal lines to illustrate the discussion.)  In particular:

{\em Mimas and Enceladus}: These moons have the weakest surface 
gravities and the smallest escape velocities of those under consideration, as well as the highest cometary impact velocities.  This combination means that an
impactor will create the largest crater on these satellites, and further,
will generate more mass for sesquinary craters than for secondaries.

{\em Tethys, Dione, and Rhea}: Larger than Mimas and Enceladus,
these moons have higher surface gravities and escape velocities.
In addition, impact speeds are lower.  On these objects, there's
generally more mass to create secondary craters than sesquinary 
craters.

{\em Iapetus}: Despite its similarity in size to Dione and Rhea,
Iapetus does not have much ejecta available for either secondaries
or sesquinaries, due to the low primary impact velocity of around
6~km/s.  The low impact velocity has two effects: the resulting
primary crater and its ejecta mass are smaller, and a slower-moving 
primary impact generates slower moving ejecta.

{\em The Galilean Satellites}: These objects are clearly distinct 
from the Saturnian
satellites, in that they have higher fractions of mass available
for secondary craters, and lower fractions of mass available for
sesquinary craters.  Their higher escape velocities mean they
retain more ejecta as secondary craters.  For Europa, the combination
of high impact velocity (which generates a large primary with lots
of fast moving ejecta) and higher surface gravity means there is
a significant amount of mass available to make secondary craters.
This explains why clusters of secondary
craters are everywhere on Europa \citep{bierhaus2005} -- even 
though that moon has a
relatively low density of primary craters -- and yet clusters are
rare to absent on the much more heavily cratered Saturnian
satellites. 

\section{Cassini Image Data}

We now turn to our analysis of Cassini images of three of Saturn's mid-sized satellites and our techniques for measuring craters on the moons.

\subsection{Description of Image Processing}

We obtained the image data from the NASA Planetary Data System (PDS) website ({\tt http://img.pds.nasa.gov/}), and performed all image processing using the USGS planetary image software package
ISIS version 3, which provides several routines to manage Cassini ISS
data.  We used a combination of the Cassini-ISS-specific
routines and additional ISIS routines to generate image mosaics
that were the basis for our measurements.  The processing can be
considered to consist of two phases: the first phase is ``image preparation",
which is a sequence of steps that generates a properly calibrated
image; the second phase is ``mosaic preparation", which collates
information from each calibrated image to generate the mosaic. [For more information on each routine,
see the ISIS website ({\tt http://isis.astrogeology.usgs.gov/Application/})].

The following is a brief summary of the processing steps for 
the image preparation phase.

\begin{figure}
\begin{center}
\epsfxsize=100mm
\includegraphics[width=5truein]{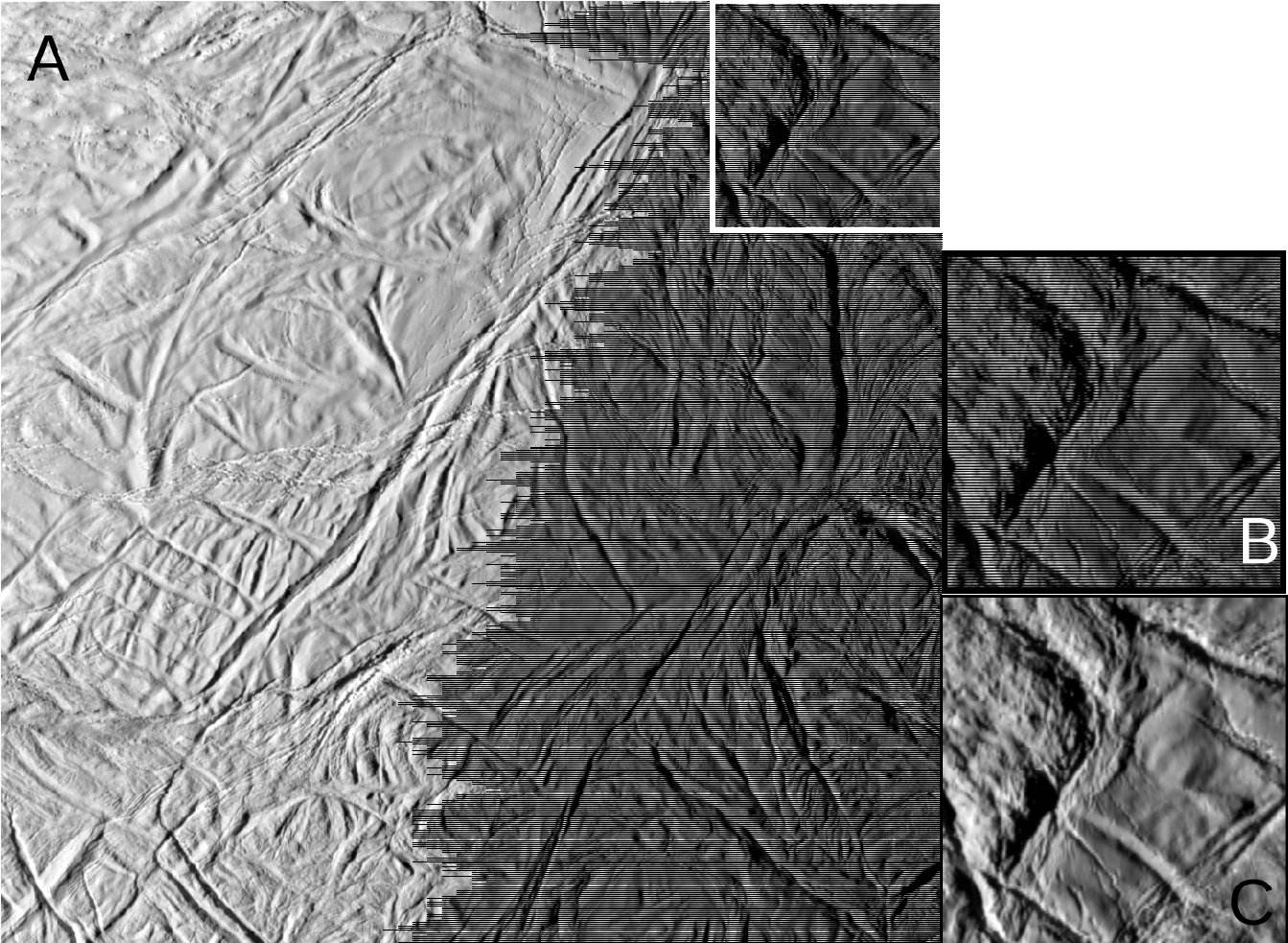}
\caption[Gap-filling a Cassini ISS image.]
{(A) is the full frame, unprocessed Cassini image N1500062262\_1,
which is one of the images of the ISS\_011EN\_MORPH002\_PRIME mosaic.  
(See Figure~\ref{fig:encelmorh002}.)  The image is complete
on the left-hand side, but every-other line on the right-hand side
ends prematurely due to the data compression algorithm.  The termination
point of every-other line is different for every line pair, because 
different scene
content within each line pair will compress with different efficiencies.
The white box outlined in the upper-right is a $250 \times 250$~pixel
portion of the image.  (B) is simply a slightly magnified portion
of the region within the $250 \times 250$ white box, better illustrating 
the every-other-line
data gaps.  (C) is the gap-filled version of (B), accomplished by using
the ISIS routine {\em lowpass}.  Essentially, each blank pixel is
populated by an intensity value that is the average of the intensity
values in the pixels immediately above and below the blank pixel.}
\label{fig:lowpass}
\end{center}
\end{figure}

\begin{enumerate}

\item {\em ciss2isis}:  imports a Cassini ISS image into ISIS by collating the separate data
file (.IMG) and label file (.LBL) from PDS into a single
ISIS-format cube file (.cub).  This is a Cassini-specific routine.

\item {\em cisscal}: performs radiometric corrections
to the images.  This is a Cassini-specific routine \citep{west2010}. 

\item {\em spiceinit}: attaches geometry information for the spacecraft,
instrument, and target body to determine the geometric properties 
(e.g., resolution, latitude, longitude) of each pixel in the image.

\item {\em trim}: removes a user-specified number of
pixels from a user-specified edge of the image.  Because ISS images
often have a one or two pixel border of invalid data around an image,
we removed a two-pixel wide border from each edge of an image.

\item {\em lowpass}: applies a $N \times M$ moving filter to the image.  ISS
image data are often returned using a compression scheme that assigns
a fixed data volume per two-line pair.  The data volume is more than
sufficient for a single line, but not enough for the entire second
line.  The completeness of the second line depends on the scene content 
(and thus the ability of the algorithm to compress the data), but is
never total.  
This results in horizontal variably sized ``jail bars", i.e., black bars 
with no image data, every other line in the image.  To eliminate these
data gaps, we used a $1 \times 3$ (line by sample) moving filter to generate
image data.  In other words, each data-gap pixel became an image pixel,
with an intensity value that is the average of the pixels above and 
below it.  See Figure~\ref{fig:lowpass}.

\item {\em camstats}: outputs a text file containing summary
information on the latitude, longitude, resolution, phase angle,
and several other image properties.  While {\em camstats} does no image processing,
it is very useful to have a text-searchable summary for
each image.

\end{enumerate}

Once the previous steps have been applied to a set of images that
comprise a mosaic, one must collect information on each image to
determine the properties of the mosaic.
The following is a summary of the routines and other steps used for the 
mosaic preparation phase.

\begin{enumerate}
\item {\em mosrange}: computes and outputs the latitude and longitude
extent, and pixel resolution, of a set of images.  The results are
output to a ``.map" file.  The .map file contains the 
required latitude and longitude extent of the mosaic, the mosaic
projection type, and other details necessary to reproject and combine the
individual images into the mosaic.

\item {\em qnet}: creates and edits a control network. This step 
requires a significant amount
of user interaction.  Briefly, the user loads the set of images
for the mosaic into
{\em qnet}, which is a graphical user interface that allows the 
user to identify ``tie points" between
images.  A tie point is a common feature between two (or more)
images.  Although the geometry information loaded into each image
during the {\em spiceinit} step (mentioned above) contains roughly correct
geometry information to align neighboring images, there are still
residual errors at the pixel-level scale.  The {\em qnet} routine enables
one to identify the residual misalignment between the
SPICE files (accessed using {\em spiceinit}) and the actual image
alignment.
An output of the user's interaction with {\em qnet} is a network file
(a ``.net" file), which contains the corrections to the image alignments.

\item {\em jigsaw}: more precisely aligns the images to sub-pixel 
levels.  The inputs to jigsaw include the set of images to be
mosaiced, and the network file created using {\em qnet}.

\item {\em cam2map}: using the data from the ``.map" file, this
routine converts each image from the 2D plane of the
image detector to a map-projected image.  ISIS provides several
map projection options; we typically used sinusoidal (Mercator equal-area)
projections to generate mosaics for crater measurement.

\item {\em automos}: creates a mosaic from a set of map-projected
images.  The output of {\em automos} is the product we used for
crater measurement.

\end{enumerate}

Some image mosaics consist of images acquired from rapidly
varying viewing geometry (i.e., during a close flyby), or
during an extended viewing sequence.  For these mosaics, the
native image resolution and/or phase angle (important for
identifying craters, or distinguishing them from other
features) can vary by many tens of percent, or even more
than a factor of two, between the first and last images of
the mosaic sequence.  In these cases, we do not use all
images that belong to the complete sequence; instead, we
use those images whose resolutions differ by no more than 50\%. 
We make this choice because the ISIS routine {\em automos} creates an image 
mosaic with a single resolution (specified by the .map
file generated by the {\em mosrange} routine); it does not preserve the 
native resolution of the individual images.  Minimizing the
resolution difference between images used to generate a mosaic
ensures a relatively common completeness limit across the mosaic.

\section{Measured Crater Size-Frequency Distributions}


We display the crater size-frequency distributions (SFDs) as ``R-plots" 
[relative plots]. \citep{crater1979} Recall that SFDs can often be 
approximated as power laws, or a series of power laws, such that 
$dN/dD =k D^{-q}$, where $dN$ is the number of craters per unit 
area with diameters between $D$ and $D + dD$, and $k$ and $q$ are  positive 
constants. For example, primary craters on the terrestrial planets 
follow distributions with $q \sim 3$. To enhance structure, the 
R-plot divides $dN/dD$ by a power law with $q = 3$, i.e.:
\begin{equation}
R_v = \frac{dN}{dD}\frac{1}{D^{-3}}
\end{equation}
At diameters less than a few km, terrestrial 
planet SFDs are roughly horizontal lines in 
R-plots, which is to say, $q \sim 3$. 
For subsequent discussion, we use the following terminology:
if $R$ increases with increasing diameter, the SFD is shallow, i.e., 
dominated by large craters. For example, if $R \propto D$, 
$dN/dD \propto D^{-2}$.

\subsection{Enceladus}

We focused our initial measurements on the young terrain of Enceladus,
because those regions should be the best representation of the true
primary crater population, and the least affected by secondary and/or
sesquinary craters.  We also measured some regions of heavily cratered
terrain to compare the crater populations between young and old(er)
regions of Enceladus.  Table~\ref{tbl:enceladus} summarizes the
regions we've measured, and discussion of the measurements follows.

\begin{figure}
\begin{center}
\epsfxsize=100mm
\includegraphics[width=5truein]{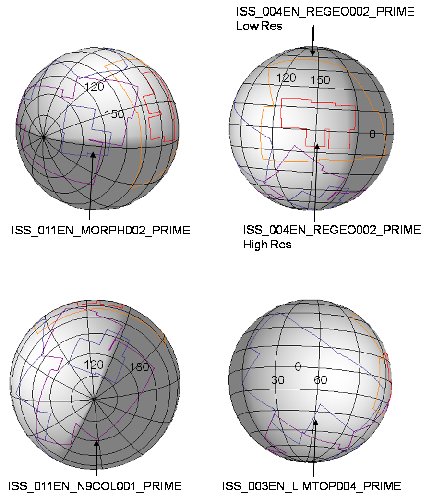}
\caption[Outlines of measured Enceladus mosaics.]
{The outlines of the Enceladus mosaics we measured.  The latitude lines are in
increments of $15^{\circ}$, and the longitude lines are in increments
of $30^{\circ}$ and are in positive-east longitude.  The measured
regions span latitudes
from the south pole to a little north of the equator,  and span
a variety of longitudes.  The shadowing on the sphere is arbitrary,
and does not reflect the terminator in any of the measured image
sequences.}
\label{fig:enceloutlines}
\end{center}
\end{figure}

{\em Young Terrain}: We measured the crater populations within several
regions of the young terrains of Enceladus, including portions of
the \\ ISS\_003EN\_LIMTOP004\_PRIME, ISS\_004EN\_REGEO002\_PRIME, \\
ISS\_011EN\_MORPH002\_PRIME, and ISS\_011EN\_N9COL001\_PRIME mosaics
(see Table~\ref{tbl:enceladus}).  
These mosaics span significant
areas at and near the south pole, and a variety of longitudes
(see Figure~\ref{fig:enceloutlines}).

The SFDs on young terrains  on Enceladus show a few notable features (see 
Figures~\ref{fig:encelregeo002rplot}, \ref{fig:encelregeo002loresrplot}, \ref{fig:encellimtop004rplot}, \ref{fig:encelmorh002rplot}, and \ref{fig:enceln9col001rplot}).
First, they all display a differential slope of roughly -2 ($q \sim 2$), regardless of location.  Second, the crater density roughly correlates with latitude: the southernmost
regions have the lowest crater densities, while the lower latitudes
have higher crater densities.  Third is the ``noise" in the crater
SFDs; although the average trend for each region is a -2 differential
slope, there are density excursions (``bumps and wiggles") that
depart from the linear trend.  We have not yet ascertained what
causes these departures, but an initial explanation is that
these density variations are evidence for a dispersed and
generally unclustered secondary crater population.

{\em Old(er) Terrain}: There are two key features of the crater
SFD measured on the older, more heavily cratered terrain.  First,
at diameters larger than a few km, the SFD has a
-3 differential slope.  Second, at diameters less than
a few km, there is a clear transition in the SFD to a roughly
-2 differential slope that parallels the SFD seen on the
young terrains of Enceladus.

The absence of a steeply-sloped SFD (i.e., $q > 3$) at small crater diameters,
even within heavily cratered regions,
supports our prediction that secondaries should have minimal
influence on the small crater populations of Enceladus.  The
presence of a steeper SFD at larger diameters is an
intriguing evolution of what we preliminarily predict to
be a -2 differential production population.  We raise two
points to address this evolution.

First, if any secondaries form on Enceladus, we expect they will be 
only the slowest moving, and therefore the largest.  (Just as there
exists a clear inverse mass-velocity relationship for the crater
ejection process as a whole, there is also a more granular
inverse mass-velocity relationship for the fragments themselves:
higher-velocity fragments are smaller, while slower-velocity
fragments are larger.  This is why secondary craters trend to smaller
diameters with increasing distance from their primary crater, even
though the fragment ejection/impact velocity is growing.  (See
\cite{mcewen2006} for more discussion on this in regards to secondaries, or
\cite{bart2010} in regards to ejected boulders.)
In addition, these slow
moving fragments will not re-impact near their parent primary,
as discussed in Section~\ref{subsec:nearfar}.  Rather, they will travel some
significant distance away, and may be unrecognizable as
secondaries due to lack of the distinct morphologies of
tightly packed secondary craters.  The end result is a
population of craters that: ({\em i}) follow a steeper SFD but are
not clearly secondaries surrounding a certain primary; and
({\em ii}) extend down to some minimum diameter but
no smaller (i.e., any smaller fragments ejected at higher
velocities will escape).

The second point is that crater measurements by \citet{kirchoff2009},
which cover a larger region of heavily
cratered terrains, show a similar behavior, namely a
decreasing crater spatial density 
below diameters of a few km,
and variations in density at larger crater diameters.
Some cratered regions are at saturation densities, while
others are not.  It is possible that a combination of
saturation effects, and a production function that is
more steeply sloped at larger diameters, leads to the
observed crater distributions.

\subsection{The Non-Detection of Craters at the South Pole}
The images provide essentially complete longitudinal coverage of
the south pole between latitudes of  -75$^\circ$ S and -90$^\circ$ S.  
We did not identify a
single crater in this region at the limiting image scale of
$\sim 123$~m/pixel, or a completeness limit of just over 600~m (i.e., $\sim 5$~pixels).
Based on the lack of observed craters in this region, we set
an upper limit on crater density.  The area of a spherical
cap is $A_c = 2 \pi R_m^2 (1 - \sin \phi)$, where $R_m$ is the 
moon's radius and $\phi$ is the latitude; thus the area of
the region southward of -75$^\circ$ S is $13,607 \mathrm{km}^2$.

If $n$ is the true crater density (number of craters
per square km) and $A_c$ is the surface area, then
the expected number of craters is $<\!N\!> = n A_C$.
Because primary cratering is a Poisson random process,
the probability that the observed
number of events equals $k$ is:

\begin{equation}
P(N=k) = \frac{e^{-\mu} \mu^k}{k!}
\end{equation}
\noindent
where $\mu$ is the expected value.  In our case, $\mu=n A_c$,
and $k=0$.  Thus the probability becomes:

\begin{equation}
P(N=0)= \frac{e^{-n A_c} (n A_c)^0}{0!} = e^{-n A_c}
\end{equation}
For a 99\% confidence level, or $P=0.01$, we derive an upper limit on the 
true crater density of $n=3.38 \times 10^{-4}$~craters/$\mathrm{km}^2$.
For a diameter bin size from $D_{\rm min} = 600$~m to
$D_{\rm max}=600 \sqrt{2} \sim 849$~m, and bin ``middle" of
$D_{\rm mid}=683$~m, the corresponding upper limit on the R-value for the
south pole of Enceladus is:

\begin{equation}
R_{sp}=\frac{n}{D_{\rm max}-D_{\rm min}} \frac{1}{D_{\rm mid}^{-3}} = 4.3 \times 10^{-4}
\end{equation}

This value resides within the range of densities measured in
the other young terrains.

\subsection{The Primary Crater Size-Frequency Distribution}

The primary crater SFD on the young surface of 
Europa \citep{bierhaus2009,schenk2004c} displays
an approximately -2 differential slope at crater diameters between
a few km and a few tens of km, as do younger terrains
on Ganymede (bright terrain and floors/ejecta blankets of impact basins,
\citet{schenk2004c}) and Callisto (floors/ejecta blankets of impact basins, 
\citet{schenk2004c}).  The behavior of the primary crater SFD at
smaller diameters is masked by the extensive secondary populations
\citep{bierhaus2005}.  Thus Enceladus's young terrains provide
an important test of the primary crater SFD in the outer solar
system at diameters smaller than a few km.

The crater SFDs on the young terrains of Enceladus have a varying 
but consistent trend
of an approximately -2 differential slope.  The density varies
by up to an order of magnitude between regions; the lowest crater
density is at and near the south pole, while the higher crater densities
are at the mid- to northern latitudes.  The consistency in the
crater SFD, within young terrain and regardless of crater density,
strongly suggests that this shape is in fact the production population.

The south polar plumes deposit material on the surface, and
\citet{kirchoff2009} find morphological evidence that
some craters do have a ``softened" appearance.  However, deposition
from the plumes cannot account for the shallow small crater SFD
we see on Enceladus.  \citet{kempf2010} predict maximum deposition
rates of about 1~mm/yr, but only for locations within 100~m of
a plume.  For more distant locations, beyond 10~km, the deposition
rate drops off to less than $10^{-3}$ mm/yr, and the majority of
the area that sees plume deposits have deposition rates more
like $10^{-5}$~mm/yr.  A 500~m diameter crater with a 100~m
depth would take 1~Gyr to disappear by plume infill.  Meanwhile
500~m diameter craters, using the estimates
of \citet{zahnle2003}, are expected to form on Enceladus on 
timescales of 0.01--1~Myr, or at least three orders of magnitude 
faster.  Thus we conclude that the small crater
SFD on Enceladus young terrains (largely) reflects the production population.

The Enceladus young terrain crater SFD, combined with observations from 
the Galilean satellites, provide preliminary cross-planetary evidence for
a $\sim -2$ differential-slope, primary-crater population at crater
diameters less than a few tens of km in the outer Solar System.  This argues for an impacting
population (i.e., comets) that have a similar SFD for projectile 
diameters in the range from $<100$~m to several km.

\subsection{Mimas}

We measured an image mosaic that includes the large ($\sim 140$~km)
Herschel crater and surrounding region.  Table~\ref{tbl:mimas} lists
the images, and Figures~\ref{fig:mimasimage} and \ref{fig:mimasrplot}
show the mosaic and R-plot, respectively.  The measurements are for
the region outside Herschel crater.  Much like the measurements on
the heavily cratered terrain of Enceladus, the crater SFD for
diameters larger than 5~km has a roughly -3 differential slope
(which is flat on an R-plot).  At smaller diameters, the crater
density decreases, and the SFD transitions to a roughly -2 differential
slope. 

Because the measurements are in the region immediately outside
the Herschel crater, it is not unreasonable to expect that the
large crater has affected the local small crater population.
However, we note that \citet{kirchoff2010} measured a
large region of Mimas that includes areas not adjacent to Herschel,
and find a very similar SFD (compare their Figure~2 with our
Figure~\ref{fig:mimasrplot}).  

\subsection{Rhea}

We measured the crater SFD in images located near the leading face
of Rhea.  Table~\ref{tbl:rhea} lists the images used to make the
mosaic seen in Figure~\ref{fig:rheaimage}, and Figure~\ref{fig:rhearplot}
is the R-plot of our data.  We measured almost 7500 craters in just the
four images outlined in Figure~\ref{fig:rheaimage}.

Unlike Mimas and Enceladus, the crater density does not decrease
at diameters smaller than a few km; indeed, Figure~\ref{fig:rhearplot}
shows an increase in crater density at smaller diameters.  This
observation matches the prediction of Section~\ref{sec:calcs}, namely that
Rhea (along with Tethys and Dione) is sufficiently massive that it
can retain a measurable secondary crater population.  This is 
explicitly illustrated by an image sequence on Rhea that captures (one of
the few examples imaged to date on Saturnian satellites) 
a secondary crater cluster near the 48-km ray crater Inktomi \citep{wagner2011}.

\section{Sesquinaries: Where Art Thou?}

The correlation between predicted and observed secondary crater
populations on the Saturnian moons is encouraging.  What remains
puzzling is the absence of a measurable signature of sesquinary
craters on Enceladus, and especially on Mimas.  The fragments
that make secondary craters follow an inverse mass-velocity
relationship (e.g., \citet{mcewen2006}), which partly explains
their generally steep SFDs.  Because sesquinary fragments from
moons with low escape velocities are presumably the same fragments
that make secondary craters on moons with higher escape velocities, 
one would logically expect that Mimas, with its multitude of
large craters, would be covered in a steeply-sloped population of small
craters made from sesquinary fragments.  Yet the crater SFD
clearly show a decrease in crater density at smaller sizes on an R-plot.
Some of the small craters may, in fact, be sesquinaries, and at the
moment we have no means to identify whether a small crater is
primary or sesquinary.  But the fact remains that the amount
of escaped ejecta (see Table~\ref{tbl:impmass}) per impact is
factors of several greater than the impactor mass.  Given that
the sesquinary fragments mostly re-impact their source moon
\citep{alvarellos2005}, there should be a high-density small
crater population made from sesquinary fragments.
What happens to those fragments?

\citet{zahnle2008} estimate
that Ionian sesquinaries on Europa should dominate the small
crater population on Europa, provided that Melosh's spall model
can be applied without modification to the 2.5~km/s ejection
velocities required for spalls to escape Io.  Yet observations 
\citep{bierhaus2005} clearly show that most small craters on
Europa are Europan secondaries.  \citet{zahnle2008}
wondered whether the Melosh (e.g., \citet{melosh1984}) spall model
applied to the higher velocity fragments that are likely the source
of the sesquinary population.  The spall model predicts a plate-like
geometry for the fragments, i.e.\ they are much larger in length and
width than they are in height.  As \citet{zahnle2008} discussed, it
may be unreasonable that such fragments survive in that state; more
reasonably, they may break up into smaller pieces whose mean radius
is approximated by the plate thickness, or by some smaller size
scale intrinsic to the target material (for Io this might be the
thickness of individual lava flows, presumably of order a meter).

\citet{schultz1985} conducted
a number of experiments that explored the morphology
of craters made by clusters of fragments rather than
individual fragments.  The craters made  by fragments
that were still relatively compact appeared similar
to traditional impact craters, albeit shallower (as
secondaries tend to be).  More dispersed fragment clusters
made craters whose morphology departed from a traditional
impact crater.

An explanation for the missing sesquinary craters may lie in
the mechanical construction of the ejecta fragments, and the
greater (much greater, in many cases) time between launch
and impact for sesquinary craters and secondary craters.
Adjacent secondary craters are made by fragments with short
flight times, less than a few minutes and sometimes less than a
minute.  Distant secondary craters are made by fragments with
flight times of up to tens of minutes, depending on the moon.
However, \citet{alvarellos2005} show that sesquinary fragments in
the Saturnian system typically have lifetimes of decades
 -- some even survive for 10,000 years. This
is between thousands to one million orbital periods of the moons, and thus
the fragments have orders of magnitude more time to disperse
than do fragments that make secondary craters. 
Close encounters with the parent moon may provide
additional dispersive pulses to weakly bound fragments.
The smallest discrete fragment, resistant to any further
disruption or dispersion, finally impacts its source moon;
perhaps its size is sufficiently diminutive that the
resulting crater is below the imaging resolution of the
current Cassini ISS data. For example, a 300-m crater on Enceladus, 
which we would probably see in the highest-resolution mosaics 
(Table 4), would be created by an 85-m diameter fragment striking 
at the moon's escape velocity, assuming a $45^\circ$ impact angle 
(Eq.~\ref{eqn:craterrad}). For a km-sized crater on Enceladus, 
which we would see in any of the mosaics, the corresponding impactor 
diameter is 380~m. These impactor sizes are intermediate between the 
characteristic sizes for rubble and spalls calculated by 
\citet{alvarellos2005} (see their Figures~2 and 4), so it seems 
plausible that modest fragmentation in orbit could render the 
sesquinaries unobservably small. See \citet{alvarellos2008} and 
\citet{zahnle2008} for further discussion of fragment sizes.

\section{Summary}

The following are the key observations and outcomes of this analysis:

\begin{enumerate}

\item The mass available to make secondary craters depends upon the relative magnitudes of $v_{\rm min}$
(the minimum velocity required to make a secondary crater) and the target moon's modified escape velocity
$v^H_{\rm esc}$, in addition to impactor and target mechanical properties (e.g., material
strengths and density contrasts).  Objects for which 
$v^H_{\rm esc} < v_{\rm min}$ should have no
secondary crater population.  For the current discussion, that regime
includes Mimas and perhaps Enceladus (depending upon the appropriate
value of $v_{\rm min}$ for Enceladus -- some variability in the small
crater SFD seen on Enceladus' young terrains provides preliminary evidence
that secondaries may in fact form on Enceladus).  More broadly, this applies to
all small moons and most minor planets.

\item  A low surface gravity also means that primary craters will lack
an adjacent secondary crater population,
as even low-velocity ejecta will travel far from the parent
primary, perhaps even significant fractions of the body's circumference.

\item Rhea, Tethys, and Dione are sufficiently
massive that we expect them to retain measurable secondary crater populations.  Our 
measurements of Rhea (as well as those of \citet{kirchoff2010})
support this prediction.  However, these moons will lack dense
populations of adjacent secondaries due to the process described above
in (2).

\item Iapetus should not have significant secondary or
sesquinary crater populations.  This is a consequence of the
low primary impact velocities due to Iapetus' large distance from Saturn, which
overall generate smaller primary craters with less
ejecta, and also generate slower-moving ejecta.

\item The Galilean satellites, and Europa in particular, are in
the ``sweet spot" for secondary crater production.  High impact
velocities generate large craters with lots of fast-moving ejecta,
and the moons are massive enough to retain that ejecta to form
secondaries.  The same is true for the Moon, Mars, and Mercury.

\item The crater SFD measured on young terrains of Enceladus generally
follows a -2 differential slope.
The similarity between this SFD on Enceladus, the primary crater
SFD on the young jovian moon Europa, and recent observations of young
rayed craters on other Saturnian satellites \citep{schenk2011} 
provide growing evidence that this is, in fact, the production SFD at
craters of these sizes,
and that comets making these craters (comets less than a few hundred meters
in diameter) follow a similar distribution. 
A slope of -2 for the nuclei of Jupiter-family comets (JFCs) was recently published by \citet{weiler2011}, but this result cannot be taken as a confirmation of our work because (1) there are very few determinations of nuclear sizes for sub-km comets; (2) few JFCs have been discovered by surveys with well-characterized observational biases; and (3) since JFCs have, by definition, been active at some point, they have undergone physical evolution, which will tend to flatten their size distribution. Triton appears to have a significantly steeper crater SFD \citep{stern2000, mckinnon2010}, but its huge apex-antapex asymmetry seems inconsistent with heliocentric impactors \citep{schenk2007}, thus making the link between crater sizes and the SFD of the impactors a thorny one.

\item The fate of sesquinary ejecta remains elusive.  There is little
doubt that large impacts on small satellites send significant amounts
of mass into orbit, mass that would otherwise make secondary craters on
larger moons.  However, unlike secondaries -- which leave a distinct
imprint on small crater SFDs by adding steep slopes -- sesquinaries
do not express themselves so ostentatiously.  Heavily cratered regions
on Enceladus, and
especially Mimas, should have small crater populations near or at
saturation density due to the effect of sesquinaries, and yet both moons
display a distinct decrease in small crater density.  One explanation
may be that the fragments responsible for secondary craters are not
yet sub-divided into their smallest discrete components.  The
significantly longer time spent as sesquinary fragments allows
the ejecta fragments to further separate into their smallest discrete
components, which form craters too small to be resolved by current
image data. On the other hand, Cassini discovered three small moons - Pallene, Methone, and Anthe - with diameters of $\sim 5$, 3, and 2~km, respectively, orbiting between Mimas and Enceladus. \citet{alvarellos2005} speculated that these moons might be spalls launched in an event like the Herschel impact. Thus there is some reason to believe that large fragments can indeed be launched (or accrete in orbit around Saturn) and survive for long periods of time.

\item The bulk of the crater SFDs across the Saturnian satellites
{\em may} be explained by a single impacting population.  The
variation in impact velocity and surface gravities across the moons
means that a single impacting population will generate different
primary crater SFDs on each satellite.  For a given sized impactor,
the variation in primary crater size is followed by variation in ejecta mass and
ejecta speeds available to make secondary and (probably) sesquinary
craters.  For example, on Iapetus a 1~km comet makes an approximately 8~km
transient crater with $\sim 2 \times 10^{11}$~kg available to make secondaries,
while the same impactor makes a 17~km transient crater on Mimas, with
no mass available to make secondaries.  The mass available to
make secondaries will travel different distances across the moons.
The superposition of the varying primary crater, and resulting
secondary (and sesquinary) crater distributions may explain the
crater SFDs seen on the satellites.

\item Finally, we propose an update to the Voyager-era interpretation
of the Saturnian cratering record: Population I is likely dominated
by heliocentric comets, and appears on all Saturnian satellites as
the source of craters above several km diameter; Population II is a
result of secondary (and perhaps sesquinary) craters, with significantly
varying signatures between the satellites, due to differences in
primary impact velocities, surface gravities, and escape speeds.

\end{enumerate}

\section{Acknowledgements}
Michelle Kirchoff kindly provided detailed discussions of her
measurements.
We thank the Cassini Data Analysis Program for supporting this research.

\label{lastpage}

\bibliography{bibv4}

\begin{thebibliography}{61}
\providecommand{\natexlab}[1]{#1}
\providecommand{\url}[1]{\texttt{#1}}
\expandafter\ifx\csname urlstyle\endcsname\relax
  \providecommand{\doi}[1]{doi: #1}\else
  \providecommand{\doi}{doi: \begingroup \urlstyle{rm}\Url}\fi

\bibitem[{Alvarellos} et~al.(2002){Alvarellos}, {Zahnle}, {Dobrovolskis}, and
  {Hamill}]{alvarellos2002}
J.~L. {Alvarellos}, K.~J. {Zahnle}, A.~R. {Dobrovolskis}, and P.~{Hamill}.
\newblock {Orbital Evolution of Impact Ejecta from Ganymede}.
\newblock \emph{Icarus}, 160:\penalty0 108--123, November 2002.
\newblock \doi{10.1006/icar.2002.6950}.

\bibitem[{Alvarellos} et~al.(2005){Alvarellos}, {Zahnle}, {Dobrovolskis}, and
  {Hamill}]{alvarellos2005}
J.~L. {Alvarellos}, K.~J. {Zahnle}, A.~R. {Dobrovolskis}, and P.~{Hamill}.
\newblock {Fates of satellite ejecta in the Saturn system}.
\newblock \emph{Icarus}, 178:\penalty0 104--123, November 2005.
\newblock \doi{10.1016/j.icarus.2005.04.017}.

\bibitem[{Alvarellos} et~al.(2008){Alvarellos}, {Zahnle}, {Dobrovolskis}, and
  {Hamill}]{alvarellos2008}
J.~L. {Alvarellos}, K.~J. {Zahnle}, A.~R. {Dobrovolskis}, and P.~{Hamill}.
\newblock {Transfer of mass from Io to Europa and beyond due to cometary
  impacts}.
\newblock \emph{Icarus}, 194:\penalty0 636--646, April 2008.
\newblock \doi{10.1016/j.icarus.2007.09.025}.

\bibitem[{Bart} and {Melosh}(2010)]{bart2010}
G.~D. {Bart} and H.~J. {Melosh}.
\newblock {Distributions of boulders ejected from lunar craters}.
\newblock \emph{Icarus}, 209:\penalty0 337--357, October 2010.
\newblock \doi{10.1016/j.icarus.2010.05.023}.

\bibitem[{Barucci} et~al.(2008){Barucci}, {Boehnhardt}, {Cruikshank},
  {Morbidelli}, and {Dotson}]{barucci2008}
M.~A. {Barucci}, H.~{Boehnhardt}, D.~P. {Cruikshank}, A.~{Morbidelli}, and
  R.~{Dotson}.
\newblock \emph{{The Solar System Beyond Neptune}}.
\newblock 2008.

\bibitem[{Bierhaus} et~al.(2005){Bierhaus}, {Chapman}, and
  {Merline}]{bierhaus2005}
E.~B. {Bierhaus}, C.~R. {Chapman}, and W.~J. {Merline}.
\newblock {Secondary craters on Europa and implications for cratered surfaces}.
\newblock \emph{Nature}, 437:\penalty0 1125--1127, October 2005.
\newblock \doi{10.1038/nature04069}.

\bibitem[{Bierhaus} et~al.(2009){Bierhaus}, {Zahnle}, and
  {Chapman}]{bierhaus2009}
E.~B. {Bierhaus}, K.~J. {Zahnle}, and C.~R. {Chapman}.
\newblock \emph{{Europa}}, pages 161--180.
\newblock University of Arizona Press, 2009.

\bibitem[{Bottke} et~al.(2010){Bottke}, {Nesvorn{\'y}}, {Vokrouhlick{\'y}}, and
  {Morbidelli}]{bottke2010}
W.~F. {Bottke}, D.~{Nesvorn{\'y}}, D.~{Vokrouhlick{\'y}}, and A.~{Morbidelli}.
\newblock {The Irregular Satellites: The Most Collisionally Evolved Populations
  in the Solar System}.
\newblock \emph{Astronomical Journal}, 139:\penalty0 994--1014, March 2010.
\newblock \doi{10.1088/0004-6256/139/3/994}.

\bibitem[{Burns} and {Gladman}(1998)]{burns1998}
J.~A. {Burns} and B.~J. {Gladman}.
\newblock {Dynamically depleted zones for Cassinis safe passage beyond Saturns
  rings}.
\newblock \emph{Planetary and Space Science}, 46:\penalty0 1401--1407, October
  1998.
\newblock \doi{10.1016/S0032-0633(97)00147-5}.

\bibitem[{Chapman} and {McKinnon}(1986)]{chapman1986}
C.~R. {Chapman} and W.~B. {McKinnon}.
\newblock \emph{{Cratering of planetary satellites}}, pages 492--580.
\newblock IAU Colloq.~77: Some Background about Satellites, 1986.

\bibitem[{Crater Analysis Techniques Working Group} et~al.(1979){Crater
  Analysis Techniques Working Group}, {Arvidson}, {Boyce}, {Chapman},
  {Cintala}, {Fulchignoni}, {Moore}, {Neukum}, {Schultz}, {Soderblom}, {Strom},
  {Woronow}, and {Young}]{crater1979}
{Crater Analysis Techniques Working Group}, R.~E. {Arvidson}, J.~{Boyce},
  C.~{Chapman}, M.~{Cintala}, M.~{Fulchignoni}, H.~{Moore}, G.~{Neukum},
  P.~{Schultz}, L.~{Soderblom}, R.~{Strom}, A.~{Woronow}, and R.~{Young}.
\newblock {Standard techniques for presentation and analysis of crater
  size-frequency data}.
\newblock \emph{Icarus}, 37:\penalty0 467--474, February 1979.
\newblock \doi{10.1016/0019-1035(79)90009-5}.

\bibitem[{Dobrovolskis} and {Lissauer}(2004)]{dobro2004}
A.~R. {Dobrovolskis} and J.~J. {Lissauer}.
\newblock {The fate of ejecta from Hyperion}.
\newblock \emph{Icarus}, 169:\penalty0 462--473, June 2004.
\newblock \doi{10.1016/j.icarus.2004.01.006}.

\bibitem[{Dones} et~al.(2009){Dones}, {Chapman}, {McKinnon}, {Melosh},
  {Kirchoff}, {Neukum}, and {Zahnle}]{dones2009}
L.~{Dones}, C.~R. {Chapman}, W.~B. {McKinnon}, H.~J. {Melosh}, M.~R.
  {Kirchoff}, G.~{Neukum}, and K.~J. {Zahnle}.
\newblock \emph{{Icy Satellites of Saturn: Impact Cratering and Age
  Determination}}, pages 613--635.
\newblock 2009.
\newblock \doi{10.1007/978-1-4020-9217-6_19}.

\bibitem[{Dundas} and {McEwen}(2007)]{dundas2007}
C.~M. {Dundas} and A.~S. {McEwen}.
\newblock {Rays and secondary craters of Tycho}.
\newblock \emph{Icarus}, 186:\penalty0 31--40, January 2007.
\newblock \doi{10.1016/j.icarus.2006.08.011}.

\bibitem[{Giese} et~al.(2008){Giese}, {Denk}, {Neukum}, {Roatsch},
  {Helfenstein}, {Thomas}, {Turtle}, {McEwen}, and {Porco}]{giese2008}
B.~{Giese}, T.~{Denk}, G.~{Neukum}, T.~{Roatsch}, P.~{Helfenstein}, P.~C.
  {Thomas}, E.~P. {Turtle}, A.~{McEwen}, and C.~C. {Porco}.
\newblock {The topography of Iapetus' leading side}.
\newblock \emph{Icarus}, 193:\penalty0 359--371, February 2008.
\newblock \doi{10.1016/j.icarus.2007.06.005}.

\bibitem[{Hartmann}(1984)]{hartmann1984}
W.~K. {Hartmann}.
\newblock {Does crater 'saturation equilibrium' occur in the solar system?}
\newblock \emph{Icarus}, 60:\penalty0 56--74, October 1984.
\newblock \doi{10.1016/0019-1035(84)90138-6}.

\bibitem[{Horedt} and {Neukum}(1984{\natexlab{a}})]{horedt1984a}
G.~P. {Horedt} and G.~{Neukum}.
\newblock {Cratering rate over the surface of a synchronous satellite}.
\newblock \emph{Icarus}, 60:\penalty0 710--717, December 1984{\natexlab{a}}.
\newblock \doi{10.1016/0019-1035(84)90175-1}.

\bibitem[{Horedt} and {Neukum}(1984{\natexlab{b}})]{horedt1984b}
G.~P. {Horedt} and G.~{Neukum}.
\newblock {Planetocentric versus heliocentric impacts in the Jovian and
  Saturnian satellite system}.
\newblock \emph{Journal of Geophysical Research}, 89:\penalty0 10405--10410,
  November 1984{\natexlab{b}}.
\newblock \doi{10.1029/JB089iB12p10405}.

\bibitem[{Housen} and {Holsapple}(2011)]{housen2011}
K.~R. {Housen} and K.~A. {Holsapple}.
\newblock {Ejecta from impact craters}.
\newblock \emph{Icarus}, 211:\penalty0 856--875, January 2011.
\newblock \doi{10.1016/j.icarus.2010.09.017}.

\bibitem[{Kempf} et~al.(2010){Kempf}, {Beckmann}, and {Schmidt}]{kempf2010}
S.~{Kempf}, U.~{Beckmann}, and J.~{Schmidt}.
\newblock {How the Enceladus dust plume feeds Saturn's E ring}.
\newblock \emph{Icarus}, 206:\penalty0 446--457, April 2010.
\newblock \doi{10.1016/j.icarus.2009.09.016}.

\bibitem[{Kirchoff} and {Schenk}(2009)]{kirchoff2009}
M.~R. {Kirchoff} and P.~{Schenk}.
\newblock {Crater modification and geologic activity in Enceladus' heavily
  cratered plains: Evidence from the impact crater distribution}.
\newblock \emph{Icarus}, 202:\penalty0 656--668, August 2009.
\newblock \doi{10.1016/j.icarus.2009.03.034}.

\bibitem[{Kirchoff} and {Schenk}(2010)]{kirchoff2010}
M.~R. {Kirchoff} and P.~{Schenk}.
\newblock {Impact cratering records of the mid-sized, icy saturnian
  satellites}.
\newblock \emph{Icarus}, 206:\penalty0 485--497, April 2010.
\newblock \doi{10.1016/j.icarus.2009.12.007}.

\bibitem[{Levison} and {Duncan}(1997)]{levison1997}
H.~F. {Levison} and M.~J. {Duncan}.
\newblock {From the Kuiper Belt to Jupiter-Family Comets: The Spatial
  Distribution of Ecliptic Comets}.
\newblock \emph{Icarus}, 127:\penalty0 13--32, May 1997.
\newblock \doi{10.1006/icar.1996.5637}.

\bibitem[{Lorenz} et~al.(2007){Lorenz}, {Wood}, {Lunine}, {Wall}, {Lopes},
  {Mitchell}, {Paganelli}, {Anderson}, {Wye}, {Tsai}, {Zebker}, and
  {Stofan}]{lorenz2007}
R.~D. {Lorenz}, C.~A. {Wood}, J.~I. {Lunine}, S.~D. {Wall}, R.~M. {Lopes},
  K.~L. {Mitchell}, F.~{Paganelli}, Y.~Z. {Anderson}, L.~{Wye}, C.~{Tsai},
  H.~{Zebker}, and E.~R. {Stofan}.
\newblock {Titan's young surface: Initial impact crater survey by Cassini RADAR
  and model comparison}.
\newblock \emph{Geophysical Research Letters}, 34:\penalty0 L07204, April 2007.
\newblock \doi{10.1029/2006GL028971}.

\bibitem[{Marchi} et~al.(2009){Marchi}, {Mottola}, {Cremonese}, {Massironi},
  and {Martellato}]{marchi2009}
S.~{Marchi}, S.~{Mottola}, G.~{Cremonese}, M.~{Massironi}, and E.~{Martellato}.
\newblock {A New Chronology for the Moon and Mercury}.
\newblock \emph{Astronomical Journal}, 137:\penalty0 4936--4948, June 2009.
\newblock \doi{10.1088/0004-6256/137/6/4936}.

\bibitem[{McEwen} and {Bierhaus}(2006)]{mcewen2006}
A.~S. {McEwen} and E.~B. {Bierhaus}.
\newblock {The Importance of Secondary Cratering to Age Constraints on
  Planetary Surfaces}.
\newblock \emph{Annual Review of Earth and Planetary Sciences}, 34:\penalty0
  535--567, May 2006.
\newblock \doi{10.1146/annurev.earth.34.031405.125018PDF:
  http://arjournals.annualreviews.org/doi/pdf/10.1146/annurev.earth.34.031405.%
125018}.

\bibitem[{McEwen} et~al.(2005){McEwen}, {Preblich}, {Turtle}, {Artemieva},
  {Golombek}, {Hurst}, {Kirk}, {Burr}, and {Christensen}]{mcewen2005}
A.~S. {McEwen}, B.~S. {Preblich}, E.~P. {Turtle}, N.~A. {Artemieva}, M.~P.
  {Golombek}, M.~{Hurst}, R.~L. {Kirk}, D.~M. {Burr}, and P.~R. {Christensen}.
\newblock {The rayed crater Zunil and interpretations of small impact craters
  on Mars}.
\newblock \emph{Icarus}, 176:\penalty0 351--381, August 2005.
\newblock \doi{10.1016/j.icarus.2005.02.009}.

\bibitem[{McKinnon} and {Singer}(2010)]{mckinnon2010}
W.~B. {McKinnon} and K.~N. {Singer}.
\newblock {Small Impact Craters on Triton: Evidence for a Turn-up in the
  Size-frequency Distribution of Small (sub-km) KBOs, and Arguments Against a
  Planetocentric Origin}.
\newblock In \emph{AAS/Division for Planetary Sciences Meeting Abstracts \#42},
  volume~42 of \emph{Bulletin of the American Astronomical Society}, pages
  984--+, October 2010.

\bibitem[{Melosh}(1984)]{melosh1984}
H.~J. {Melosh}.
\newblock {Impact ejection, spallation, and the origin of meteorites}.
\newblock \emph{Icarus}, 59:\penalty0 234--260, August 1984.
\newblock \doi{10.1016/0019-1035(84)90026-5}.

\bibitem[{Moore} et~al.(2004){Moore}, {Schenk}, {Bruesch}, {Asphaug}, and
  {McKinnon}]{moore2004}
J.~M. {Moore}, P.~M. {Schenk}, L.~S. {Bruesch}, E.~{Asphaug}, and W.~B.
  {McKinnon}.
\newblock {Large impact features on middle-sized icy satellites}.
\newblock \emph{Icarus}, 171:\penalty0 421--443, October 2004.
\newblock \doi{10.1016/j.icarus.2004.05.009}.

\bibitem[{Neish} and {Lorenz}(2011)]{neish2011}
C.~D. {Neish} and R.~D. {Lorenz}.
\newblock {Titan's global crater population: A new assessment}.
\newblock \emph{Planetary and Space Science}, 000:\penalty0 000--+, 000 2011.
\newblock \doi{doi:10.1016/j.pss.2011.02.016}.

\bibitem[{Neukum}(1985)]{neukum1985}
G.~{Neukum}.
\newblock {Cratering records of the satellites of Jupiter and Saturn}.
\newblock \emph{Advances in Space Research}, 5:\penalty0 107--116, 1985.
\newblock \doi{10.1016/0273-1177(85)90247-9}.

\bibitem[{Plescia} and {Boyce}(1985)]{plescia1985}
J.~B. {Plescia} and J.~M. {Boyce}.
\newblock {Impact cratering history of the Saturnian satellites}.
\newblock \emph{Journal of Geophysical Research}, 90:\penalty0 2029--2037,
  February 1985.
\newblock \doi{10.1029/JB090iB02p02029}.

\bibitem[{Porco} et~al.(2005){Porco}, {Baker}, {Barbara}, {Beurle}, {Brahic},
  {Burns}, {Charnoz}, {Cooper}, {Dawson}, {Del Genio}, {Denk}, {Dones},
  {Dyudina}, {Evans}, {Giese}, {Grazier}, {Helfenstein}, {Ingersoll},
  {Jacobson}, {Johnson}, {McEwen}, {Murray}, {Neukum}, {Owen}, {Perry},
  {Roatsch}, {Spitale}, {Squyres}, {Thomas}, {Tiscareno}, {Turtle}, {Vasavada},
  {Veverka}, {Wagner}, and {West}]{porco2005}
C.~C. {Porco}, E.~{Baker}, J.~{Barbara}, K.~{Beurle}, A.~{Brahic}, J.~A.
  {Burns}, S.~{Charnoz}, N.~{Cooper}, D.~D. {Dawson}, A.~D. {Del Genio},
  T.~{Denk}, L.~{Dones}, U.~{Dyudina}, M.~W. {Evans}, B.~{Giese}, K.~{Grazier},
  P.~{Helfenstein}, A.~P. {Ingersoll}, R.~A. {Jacobson}, T.~V. {Johnson},
  A.~{McEwen}, C.~D. {Murray}, G.~{Neukum}, W.~M. {Owen}, J.~{Perry},
  T.~{Roatsch}, J.~{Spitale}, S.~{Squyres}, P.~C. {Thomas}, M.~{Tiscareno},
  E.~{Turtle}, A.~R. {Vasavada}, J.~{Veverka}, R.~{Wagner}, and R.~{West}.
\newblock {Cassini Imaging Science: Initial Results on Phoebe and Iapetus}.
\newblock \emph{Science}, 307:\penalty0 1237--1242, February 2005.
\newblock \doi{10.1126/science.1107981}.

\bibitem[{Porco} et~al.(2006){Porco}, {Helfenstein}, {Thomas}, {Ingersoll},
  {Wisdom}, {West}, {Neukum}, {Denk}, {Wagner}, {Roatsch}, {Kieffer}, {Turtle},
  {McEwen}, {Johnson}, {Rathbun}, {Veverka}, {Wilson}, {Perry}, {Spitale},
  {Brahic}, {Burns}, {Del Genio}, {Dones}, {Murray}, and {Squyres}]{porco2006}
C.~C. {Porco}, P.~{Helfenstein}, P.~C. {Thomas}, A.~P. {Ingersoll},
  J.~{Wisdom}, R.~{West}, G.~{Neukum}, T.~{Denk}, R.~{Wagner}, T.~{Roatsch},
  S.~{Kieffer}, E.~{Turtle}, A.~{McEwen}, T.~V. {Johnson}, J.~{Rathbun},
  J.~{Veverka}, D.~{Wilson}, J.~{Perry}, J.~{Spitale}, A.~{Brahic}, J.~A.
  {Burns}, A.~D. {Del Genio}, L.~{Dones}, C.~D. {Murray}, and S.~{Squyres}.
\newblock {Cassini Observes the Active South Pole of Enceladus}.
\newblock \emph{Science}, 311:\penalty0 1393--1401, March 2006.
\newblock \doi{10.1126/science.1123013}.

\bibitem[{Preblich} et~al.(2007){Preblich}, {McEwen}, and
  {Studer}]{preblich2007}
B.~S. {Preblich}, A.~S. {McEwen}, and D.~M. {Studer}.
\newblock {Mapping rays and secondary craters from the Martian crater Zunil}.
\newblock \emph{Journal of Geophysical Research (Planets)}, 112:\penalty0
  E05006, May 2007.
\newblock \doi{10.1029/2006JE002817}.

\bibitem[{Richardson}(2009)]{richardson2009}
J.~E. {Richardson}.
\newblock {Cratering saturation and equilibrium: A new model looks at an old
  problem}.
\newblock \emph{Icarus}, 204:\penalty0 697--715, December 2009.
\newblock \doi{10.1016/j.icarus.2009.07.029}.

\bibitem[{Robbins} and {Hynek}(2011)]{robbins2011}
S.~J. {Robbins} and B.~M. {Hynek}.
\newblock {Distant secondary craters from Lyot crater, Mars, and implications
  for surface ages of planetary bodies}.
\newblock \emph{Geophysical Research Letters}, 38:\penalty0 L05201, March 2011.
\newblock \doi{10.1029/2010GL046450}.

\bibitem[{Schenk} and {Murphy}(2011)]{schenk2011}
P.~M. {Schenk} and S.~W. {Murphy}.
\newblock {The rayed craters of Saturn's icy satellites (including Iapetus):
  Current impactor populations and origins}.
\newblock In \emph{Lunar and Planetary Institute Science Conference Abstracts},
  volume~42 of \emph{Lunar and Planetary Institute Science Conference
  Abstracts}, pages LOL--LOL, March 2011.

\bibitem[{Schenk} and {Pappalardo}(2004)]{schenk2004}
P.~M. {Schenk} and R.~T. {Pappalardo}.
\newblock {Topographic variations in chaos on Europa: Implications for diapir
  ic formation}.
\newblock \emph{Geophysical Research Letters}, 31:\penalty0 16703--+, August
  2004.
\newblock \doi{10.1029/2004GL019978}.

\bibitem[{Schenk} and {Zahnle}(2007)]{schenk2007}
P.~M. {Schenk} and K.~{Zahnle}.
\newblock {On the negligible surface age of Triton}.
\newblock \emph{Icarus}, 192:\penalty0 135--149, December 2007.
\newblock \doi{10.1016/j.icarus.2007.07.004}.

\bibitem[{Schenk} et~al.(2004){Schenk}, {Chapman}, {Zahnle}, and
  {Moore}]{schenk2004c}
P.~M. {Schenk}, C.~R. {Chapman}, K.~{Zahnle}, and J.~M. {Moore}.
\newblock \emph{{Ages and interiors: the cratering record of the Galilean
  satellites}}, pages 427--456.
\newblock 2004.

\bibitem[{Schmidt} and {Housen}(1987)]{schmidt1987}
R.~M. {Schmidt} and K.~R. {Housen}.
\newblock {Some recent advances in the scaling of impact and explosion
  cratering}.
\newblock \emph{International Journal of Impact Engineering}, 5:\penalty0
  543--560, 1987.

\bibitem[{Schultz} and {Gault}(1985)]{schultz1985}
P.~H. {Schultz} and D.~E. {Gault}.
\newblock {Clustered impacts - Experiments and implications}.
\newblock \emph{Journal of Geophyscal Research}, 90:\penalty0 3701--3732, April
  1985.

\bibitem[{Shoemaker}(1965)]{shoemaker1965}
E.~M. {Shoemaker}.
\newblock {Preliminary Analysis of the Fine Structure of the Lunar Surface in
  Mare Cognitum}.
\newblock In {W.~N.~Hess, D.~H.~Menzel, \& J.~A.~O'Keefe}, editor, \emph{The
  Nature of the Lunar Surface}, pages 23--+, 1965.

\bibitem[{Shoemaker} and {Wolfe}(1981)]{shoemaker1981}
E.~M. {Shoemaker} and R.~F. {Wolfe}.
\newblock {Evolution of the Saturnian Satellites: The Role of Impact}.
\newblock \emph{LPI Contributions}, 428:\penalty0 1--+, 1981.

\bibitem[{Shoemaker} and {Wolfe}(1982)]{shoemaker1982}
E.~M. {Shoemaker} and R.~F. {Wolfe}.
\newblock {Cratering time scales for the Galilean satellites}.
\newblock In D.~{Morrison}, editor, \emph{Satellites of Jupiter}, pages
  277--339, 1982.

\bibitem[{Smith} et~al.(1981){Smith}, {Soderblom}, {Beebe}, {Boyce}, {Briggs},
  {Bunker}, {Collins}, {Hansen}, {Johnson}, {Mitchell}, {Terrile}, {Carr},
  {Cook}, {Cuzzi}, {Pollack}, {Danielson}, {Ingersoll}, {Davies}, {Hunt},
  {Masursky}, {Shoemaker}, {Morrison}, {Owen}, {Sagan}, {Veverka}, {Strom}, and
  {Suomi}]{smith1981}
B.~A. {Smith}, L.~{Soderblom}, R.~F. {Beebe}, J.~M. {Boyce}, G.~{Briggs},
  A.~{Bunker}, S.~A. {Collins}, C.~{Hansen}, T.~V. {Johnson}, J.~L. {Mitchell},
  R.~J. {Terrile}, M.~H. {Carr}, A.~F. {Cook}, J.~N. {Cuzzi}, J.~B. {Pollack},
  G.~E. {Danielson}, A.~P. {Ingersoll}, M.~E. {Davies}, G.~E. {Hunt},
  H.~{Masursky}, E.~M. {Shoemaker}, D.~{Morrison}, T.~{Owen}, C.~{Sagan},
  J.~{Veverka}, R.~{Strom}, and V.~E. {Suomi}.
\newblock {Encounter with Saturn - Voyager 1 imaging science results}.
\newblock \emph{Science}, 212:\penalty0 163--191, April 1981.
\newblock \doi{10.1126/science.212.4491.163}.

\bibitem[{Smith} et~al.(1982){Smith}, {Soderblom}, {Batson}, {Bridges}, {Inge},
  {Masursky}, {Shoemaker}, {Beebe}, {Boyce}, {Briggs}, {Bunker}, {Collins},
  {Hansen}, {Johnson}, {Mitchell}, {Terrile}, {Cook}, {Cuzzi}, {Pollack},
  {Danielson}, {Ingersoll}, {Davies}, {Hunt}, {Morrison}, {Owen}, {Sagan},
  {Veverka}, {Strom}, and {Suomi}]{smith1982}
B.~A. {Smith}, L.~{Soderblom}, R.~M. {Batson}, P.~M. {Bridges}, J.~L. {Inge},
  H.~{Masursky}, E.~{Shoemaker}, R.~F. {Beebe}, J.~{Boyce}, G.~{Briggs},
  A.~{Bunker}, S.~A. {Collins}, C.~{Hansen}, T.~V. {Johnson}, J.~L. {Mitchell},
  R.~J. {Terrile}, A.~F. {Cook}, J.~N. {Cuzzi}, J.~B. {Pollack}, G.~E.
  {Danielson}, A.~P. {Ingersoll}, M.~E. {Davies}, G.~E. {Hunt}, D.~{Morrison},
  T.~{Owen}, C.~{Sagan}, J.~{Veverka}, R.~{Strom}, and V.~E. {Suomi}.
\newblock {A new look at the Saturn system - The Voyager 2 images}.
\newblock \emph{Science}, 215:\penalty0 504--537, January 1982.
\newblock \doi{10.1126/science.215.4532.504}.

\bibitem[{Stern} and {McKinnon}(2000)]{stern2000}
S.~A. {Stern} and W.~B. {McKinnon}.
\newblock {Triton's Surface Age and Impactor Population Revisited in Light of
  Kuiper Belt Fluxes: Evidence for Small Kuiper Belt Objects and Recent
  Geological Activity}.
\newblock \emph{Astronomical Journal}, 119:\penalty0 945--952, February 2000.
\newblock \doi{10.1086/301207}.

\bibitem[{Strom} and {Woronow}(1982)]{strom1982}
R.~G. {Strom} and A.~{Woronow}.
\newblock {Solar System Cratering Populations}.
\newblock In \emph{Lunar and Planetary Institute Science Conference Abstracts},
  volume~13 of \emph{Lunar and Planetary Institute Science Conference
  Abstracts}, pages 782--783, March 1982.

\bibitem[{Thomas} et~al.(2007){Thomas}, {Armstrong}, {Asmar}, {Burns}, {Denk},
  {Giese}, {Helfenstein}, {Iess}, {Johnson}, {McEwen}, {Nicolaisen}, {Porco},
  {Rappaport}, {Richardson}, {Somenzi}, {Tortora}, {Turtle}, and
  {Veverka}]{thomas2007}
P.~C. {Thomas}, J.~W. {Armstrong}, S.~W. {Asmar}, J.~A. {Burns}, T.~{Denk},
  B.~{Giese}, P.~{Helfenstein}, L.~{Iess}, T.~V. {Johnson}, A.~{McEwen},
  L.~{Nicolaisen}, C.~{Porco}, N.~{Rappaport}, J.~{Richardson}, L.~{Somenzi},
  P.~{Tortora}, E.~P. {Turtle}, and J.~{Veverka}.
\newblock {Hyperion's sponge-like appearance}.
\newblock \emph{Nature}, 448:\penalty0 50--56, July 2007.
\newblock \doi{10.1038/nature05779}.

\bibitem[{Vickery}(1986)]{vickery1986}
A.~M. {Vickery}.
\newblock {Size-velocity distribution of large ejecta fragments}.
\newblock \emph{Icarus}, 67:\penalty0 224--236, August 1986.
\newblock \doi{10.1016/0019-1035(86)90105-3}.

\bibitem[{Wagner} et~al.(2011){Wagner}, {Neukum}, {Wolf}, {Schmedemann},
  {Denk}, {Stephan}, {Roatsch}, and {Porco}]{wagner2011}
R.~J. {Wagner}, G.~{Neukum}, U.~{Wolf}, N.~{Schmedemann}, T.~{Denk},
  K.~{Stephan}, T.~{Roatsch}, and C.~C. {Porco}.
\newblock {Bright ray craters on Rhea and Dione}.
\newblock In \emph{Lunar and Planetary Institute Science Conference Abstracts},
  volume~42 of \emph{Lunar and Planetary Institute Science Conference
  Abstracts}, pages LOL--LOL, March 2011.

\bibitem[{Weiler} et~al.(2011){Weiler}, {Rauer}, and {Sterken}]{weiler2011}
M.~{Weiler}, H.~{Rauer}, and C.~{Sterken}.
\newblock {Cometary nuclear magnitudes from sky survey observations}.
\newblock \emph{Icarus}, 212:\penalty0 351--366, March 2011.
\newblock \doi{10.1016/j.icarus.2010.12.026}.

\bibitem[{West} et~al.(2010){West}, {Knowles}, {Birath}, {Charnoz}, {di Nino},
  {Hedman}, {Helfenstein}, {McEwen}, {Perry}, {Porco}, {Salmon}, {Throop}, and
  {Wilson}]{west2010}
R.~{West}, B.~{Knowles}, E.~{Birath}, S.~{Charnoz}, D.~{di Nino}, M.~{Hedman},
  P.~{Helfenstein}, A.~{McEwen}, J.~{Perry}, C.~{Porco}, J.~{Salmon},
  H.~{Throop}, and D.~{Wilson}.
\newblock {In-flight calibration of the Cassini imaging science sub-system
  cameras}.
\newblock \emph{Planetary and Space Science}, 58:\penalty0 1475--1488,
  September 2010.
\newblock \doi{10.1016/j.pss.2010.07.006}.

\bibitem[{Wood} et~al.(2010){Wood}, {Lorenz}, {Kirk}, {Lopes}, {Mitchell},
  {Stofan}, and {Cassini RADAR Team}]{wood2010}
C.~A. {Wood}, R.~{Lorenz}, R.~{Kirk}, R.~{Lopes}, K.~{Mitchell}, E.~{Stofan},
  and {Cassini RADAR Team}.
\newblock {Impact craters on Titan}.
\newblock \emph{Icarus}, 206:\penalty0 334--344, March 2010.
\newblock \doi{10.1016/j.icarus.2009.08.021}.

\bibitem[{Zahnle} et~al.(1998){Zahnle}, {Dones}, and {Levison}]{zahnle1998}
K.~{Zahnle}, L.~{Dones}, and H.~F. {Levison}.
\newblock {Cratering Rates on the Galilean Satellites}.
\newblock \emph{Icarus}, 136:\penalty0 202--222, December 1998.
\newblock \doi{10.1006/icar.1998.6015}.

\bibitem[{Zahnle} et~al.(2001){Zahnle}, {Schenk}, {Sobieszczyk}, {Dones}, and
  {Levison}]{zahnle2001}
K.~{Zahnle}, P.~{Schenk}, S.~{Sobieszczyk}, L.~{Dones}, and H.~F. {Levison}.
\newblock {Differential Cratering of Synchronously Rotating Satellites by
  Ecliptic Comets}.
\newblock \emph{Icarus}, 153:\penalty0 111--129, September 2001.
\newblock \doi{10.1006/icar.2001.6668}.

\bibitem[{Zahnle} et~al.(2003){Zahnle}, {Schenk}, {Levison}, and
  {Dones}]{zahnle2003}
K.~{Zahnle}, P.~{Schenk}, H.~{Levison}, and L.~{Dones}.
\newblock {Cratering rates in the outer Solar System}.
\newblock \emph{Icarus}, 163:\penalty0 263--289, June 2003.
\newblock \doi{10.1016/S0019-1035(03)00048-4}.

\bibitem[{Zahnle} et~al.(2008){Zahnle}, {Alvarellos}, {Dobrovolskis}, and
  {Hamill}]{zahnle2008}
K.~{Zahnle}, J.~L. {Alvarellos}, A.~{Dobrovolskis}, and P.~{Hamill}.
\newblock {Secondary and sesquinary craters on Europa}.
\newblock \emph{Icarus}, 194:\penalty0 660--674, April 2008.
\newblock \doi{10.1016/j.icarus.2007.10.024}.

\end{thebibliography}
\bibliographystyle{plainnat}

\clearpage




\begin{table}
\begin{center}
\begin{tabular}{lcccc}
{\em Moon} & $v_i^1$ [km/s] & $v_{\rm esc}$ [m/s] & $v^H_{\rm esc}$ [m/s] & $g$ [m/$\mathrm{s}^2$]\\ \hline \hline
Mimas     & 27 & 159 & 130 & 0.064\\
Enceladus & 24 & 239 & 209 & 0.11\\
Tethys    & 21 & 393 & 345 & 0.145\\
Dione     & 19 & 510 & 467 & 0.231\\
Rhea      & 16 & 635 & 559 & 0.285\\
Iapetus   & 6.1 & 572 & 566 & 0.223\\
Europa    & 26 & 2000 & 1888 & 1.31\\
Ganymede  & 20 & 2741 & 2629 & 1.43\\
Callisto  & 15 & 2440 & 2382 & 1.235\\ \hline
\end{tabular}
\end{center}
\caption{Table of satellite properties used in our calculations. 
$^1$ Impact velocities from Zahnle et al.~(2003). $v^H_{\rm esc}$
is the modified escape velocity, see Equation~\ref{eqn:vhill}.}
\label{tbl:satdata}
\end{table}

\begin{table}
\begin{center}
\begin{tabular}{lcccccccc}
{\em Moon} & D [km] & $M_{\rm tot}$ & $M_{\rm blk}$ & $f_{\rm blk}$ & $M_{\rm sec}$ & $f_{\rm sec}$ & $M_{1.5}$ & $f_{1.5}$ \\ \hline \hline
Mimas & 17.3 & $17.485 \times 10^{13}$ & $17.294 \times10^{13}$ & 0.989 & 0.000 & 0.000 & $19.065 \times 10^{11}$ & 0.011\\
Enceladus & 14.9 & $11.267 \times 10^{13}$ & $11.134 \times 10^{13}$ & 0.988 & $4.850 \times 10^{11}$ & 0.004 & $8.430 \times 10^{11}$ & 0.007\\
Tethys & 13.5 & $8.307 \times 10^{13}$ & $8.196 \times 10^{13}$ & 0.987 & $7.547 \times 10^{11}$ & 0.009 & $3.498 \times 10^{11}$ & 0.004\\
Dione & 11.9 & $5.717 \times 10^{13}$ & $5.621 \times 10^{13}$ & 0.983 & $7.616 \times 10^{11}$ & 0.013 & $2.005 \times 10^{11}$ & 0.004\\
Rhea & 10.7 & $4.191 \times 10^{13}$ & $4.115 \times 10^{13}$ & 0.982 & $6.455 \times 10^{11}$ & 0.015 & $1.135 \times 10^{11}$ & 0.003\\
Iapetus & 7.8 & $1.630 \times 10^{13}$ & $1.610 \times 10^{13}$ & 0.988 & $1.685 \times 10^{11}$ & 0.010 & $0.321 \times 10^{11}$ & 0.002\\
Europa & 9.7 & $3.071 \times 10^{13}$ & $2.922 \times 10^{13}$ & 0.952 & $14.381 \times 10^{11}$ & 0.047 & $0.450 \times 10^{11}$ & 0.001\\
Ganymede & 8.6 & $2.178 \times 10^{13}$ & $2.075 \times 10^{13}$ & 0.953 & $10.128 \times 10^{11}$ & 0.047 & $0.198 \times 10^{11}$ & 0.001\\
Callisto & 8.0 & $1.712 \times 10^{13}$ & $1.643 \times 10^{13}$ & 0.959 & $6.790 \times 10^{11}$ & 0.040 & $0.153 \times 10^{11}$ & 0.001\\ \hline
\end{tabular}
\end{center}
\caption{The total and fractional masses for the primary crater
and resulting ejecta when $v_{\rm min} = 150$~m/s.  
The diameter listed in the second column
is the transient crater diameter predicted by Equation~\ref{eqn:craterrad},
and not the final crater diameter.  The subscript
$blk$ refers to the ejecta blanket, the subscript $sec$ refers
to secondary craters, and the subscript 1.5 refers to sesquinaries.
This case of $v_{\rm min}$ minimizes
$M_{\rm blk}$, and yet almost all the mass is still moving slower than
$v_{\rm min}$, i.e., most the mass is still in $M_{\rm blk}$. All masses
are in kg.  The fractional masses are
$f_{\rm blk}=M_{\rm blk}/M_{\rm tot}$, $f_{\rm sec}=M_{\rm sec}/M_{\rm tot}$, 
and $f_{1.5}=M_{1.5}/M_{\rm tot}$.}
\label{tbl:150dat}
\end{table}

\begin{table}
\begin{center}
\begin{tabular}{lcc|cc}
           & $v_{\rm min} = 150$~m/s & & $v_{\rm min} = 250$~m/s & \\
{\em Moon} & $M_{\rm sec}/m_i$ & $M_{1.5}/m_i$ & $M_{\rm sec}/m_i$ & $M_{1.5}/m_i$\\ \hline \hline
Mimas     & 0.0 & 6.1 & 0.0 & 6.1\\
Enceladus & 1.5 & 2.7 & 0.0 & 2.7\\ \hline
Tethys    & 2.4 & 1.1 & 0.6 & 1.1\\
Dione     & 2.4 & 0.6 & 0.9 & 0.6\\
Rhea      & 2.0 & 0.4 & 0.8 & 0.4\\ \hline
Iapetus   & 0.5 & 0.1 & 0.2 & 0.1\\ \hline
Europa    & 4.6 & 0.1 & 2.2 & 0.1\\
Ganymede  & 3.2 & 0.1 & 1.6 & 0.1\\
Callisto  & 2.2 & $< 0.1$ & 1.0 & $< 0.1$\\ \hline
\end{tabular}
\end{center}
\caption{$M_{\rm sec}$ and $M_{1.5}$ in terms of the impactor mass, $m_i$,
when $v_{\rm min} = 150$~m/s and $v_{\rm min} = 250$~m/s.  The horizontal lines
in the table divide the moons into groups based on the amount of ejecta
available for secondaries and sesquinaries.  See text for discussion.}
\label{tbl:impmass}
\end{table}

\begin{table}
\begin{center}
\tiny
\begin{tabular}{ccr}
Enceladus Mosaic & Image & Scale [m/pix]\\ \hline \hline
ISS\_003EN\_LIMTOP004\_PRIME & N1487300107\_1 & 148\\
                             & N1487300285\_1 & 141\\
                             & N1487300482\_1 & 133\\
                             & N1487300648\_1 & 126\\
                             & N1487300854\_1 & 118\\
                             &               \\
ISS\_004EN\_REGEO002\_PRIME & N1489047900\_2 & 166\\
                            & N1489048083\_2 & 158\\
                            & N1489048255\_2 & 151\\
                            & N1489048757\_2 & 131\\
                            & N1489048931\_2 & 124\\
                            & N1489049105\_2 & 117\\
                            &               \\
ISS\_004EN\_REGEO002\_PRIME & N1489049969\_2 & 82\\
                            & N1489050144\_2 & 75\\
                            & N1489050320\_2 & 68\\
                            & N1489050475\_2 & 62\\
                            & N1489050651\_2 & 55\\
                            &               \\
ISS\_011EN\_MORPH002\_PRIME & N1500062131\_1 & 80\\
                            & N1500062262\_1 & 73\\
                            & N1500062382\_1 & 67\\
                            &               \\
ISS\_011EN\_N9COL001\_PRIME & N1500061132\_1 & 129\\
                            & N1500061253\_1 & 123\\
                            & N1500061390\_1 & 116\\
                            & N1500061512\_1 & 110\\
                            & N1500061634\_1 & 104\\
                            & N1500061771\_1 & 97\\
                            & N1500061892\_1 & 91\\
\hline
\end{tabular}
\end{center}
\caption{The mosaics and individual images we used to make crater
SFD measurements on Enceladus.}
\label{tbl:enceladus}
\end{table}

\begin{table}
\begin{center}
\begin{tabular}{ccr}
Mimas Mosaic & Image & Scale [m/pix]\\ \hline \hline
ISS\_126MI\_GEOLOG001\_PRIME & N1644777693\_1 & 93\\
                             & N1644777828\_1 & 97\\
                             & N1644777993\_1 & 102\\
                             & N1644778141\_1 & 106\\
                             & N1644778308\_1 & 110\\
                             & N1644778455\_1 & 115\\
                             & N1644778567\_1 & 118\\
\hline
\end{tabular}
\end{center}
\caption{The images used to measure the crater size-frequency distribution on Mimas.}
\label{tbl:mimas}
\end{table}

\begin{table}
\begin{center}
\begin{tabular}{ccr}
Rhea Mosaic & Image & Scale [m/pix]\\ \hline \hline
ISS\_121RH\_REGMAP001\_PRIME & N1637518901\_1 & 147\\
                             & N1637519058\_1 & 148\\
                             & N1637519176\_1 & 148\\
                             & N1637519283\_1 & 149\\
                             & N1637519392\_1 & 151\\
                             & N1637519501\_1 & 152\\
                             & N1637519610\_1 & 154\\
                             & N1637519768\_1 & 156\\
                             & N1637519875\_1 & 159\\
\hline
\label{lasttable}
\end{tabular}
\end{center}
\caption{The list of images used to measure a representative population
of Rhea's crater  size-frequency distribution.}
\label{tbl:rhea}
\end{table}


\clearpage


\clearpage

\clearpage

\clearpage

\clearpage

\clearpage

\clearpage

\clearpage

\clearpage

\clearpage

\clearpage

\clearpage

\begin{figure}
\begin{center}
\epsfxsize=100mm
\includegraphics[width=5truein]{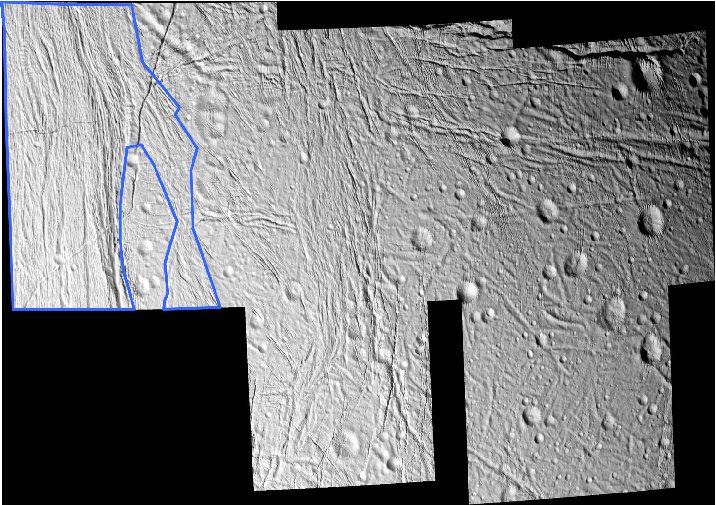}
\caption[Enceladus ISS\_004EN\_REGEO002\_PRIME.]
{Some of the higher-resolution images of the Enceladus
ISS\_004EN\_REGEO002\_PRIME mosaic.  The average
mosaic scale is 69~m/pix, and the maximum scale is 83~m/pix.
We divided the crater
measurements into the young, fractured terrain (outlined in blue) and the
heavily cratered terrain.}
\label{fig:encelregeo002}
\end{center}
\end{figure}

\clearpage

\begin{figure}
\begin{center}
\epsfxsize=100mm
\includegraphics[width=5truein]{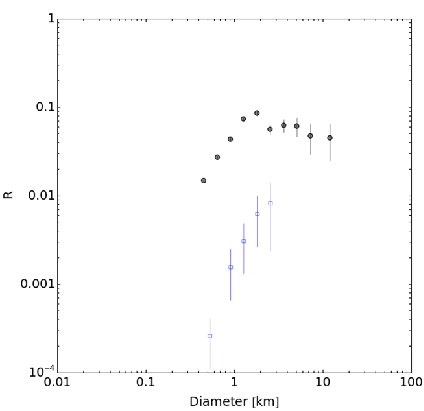}
\caption[Enceladus ISS\_004EN\_REGEO002\_PRIME.]
{The R-plot for the regions seen in Figure~\ref{fig:encelregeo002}.  The
measurements within the young,  fractured terrain are the open blue symbols,
while the measurements within the older, cratered terrain are the
filled symbols.  We measured 15 craters in the fractured terrain, 
and 2375 craters in the older terrain.  The fractured terrain displays 
a clear $\sim -2$
differential slope.  The cratered terrain has a $\sim -3$
differential slope above a few km, which transitions to a
$\sim -2$ slope at smaller diameters.  See text for discussion.}
\label{fig:encelregeo002rplot}
\end{center}
\end{figure}

\clearpage

\begin{figure}
\begin{center}
\epsfxsize=100mm
\includegraphics[width=5truein]{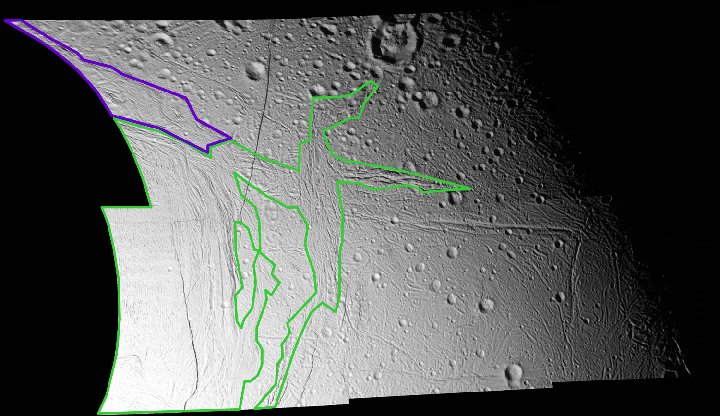}
\caption[Enceladus ISS\_004EN\_REGEO002\_PRIME lores.]
{Some of the lower-resolution images of the Enceladus
ISS\_004EN\_REGEO002\_PRIME mosaic.  The average
mosaic scale is 132~m/pix, and the maximum scale is 173~m/pix.
We divided the crater
measurements into two regions of young,  fractured terrain
(outlined in purple and green) and the
heavily cratered terrain.}
\label{fig:encelregeo002lores}
\end{center}
\end{figure}

\clearpage

\begin{figure}
\begin{center}
\epsfxsize=100mm
\includegraphics[width=5truein]{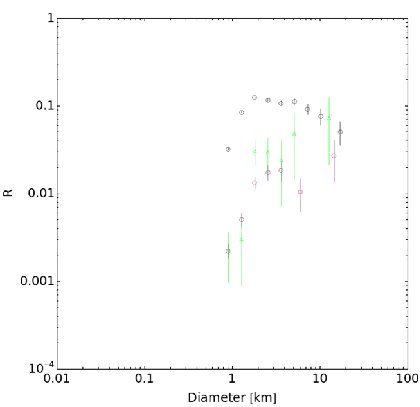}
\caption[Enceladus ISS\_004EN\_REGEO002\_PRIME.]
{The R-plot for the regions seen in Figure~\ref{fig:encelregeo002lores}.  The
measurements within the young, fractured terrain are the purple and green 
symbols, while the measurements within the older, cratered terrain are the
black symbols.  We measured 153 craters in the fractured region outlined in 
purple, 26 craters in the fractured region outlined in green, and
5181 craters in the cratered region.
Though the $R$ values show a less clear linear trend than the young 
terrain measurements
seen in Figure~{\ref{fig:encelregeo002rplot}}, the young terrain SFDs
here too  follow a $\sim -2$
differential slope.  The cratered terrain has a $\sim -3$
differential slope above a few km, which transitions to a
$\sim -2$ slope at smaller diameters.  See text for discussion.}
\label{fig:encelregeo002loresrplot}
\end{center}
\end{figure}

\clearpage

\begin{figure}
\begin{center}
\epsfxsize=100mm
\includegraphics[width=5truein]{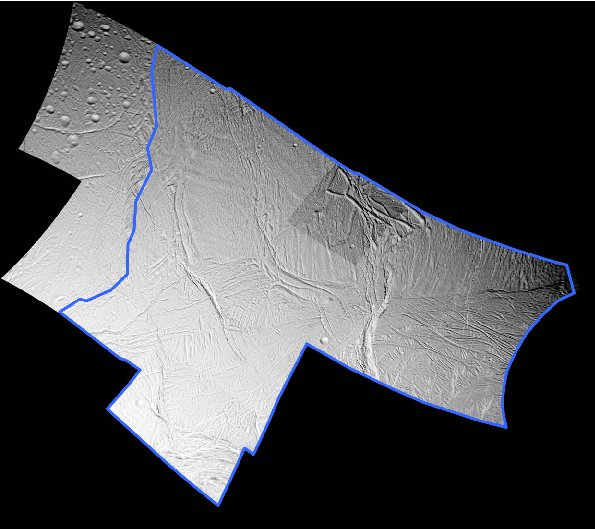}
\caption[Enceladus ISS\_003EN\_LIMTOP004\_PRIME.]
{Some of the images of the Enceladus
ISS\_003EN\_LIMTOP004\_PRIME mosaic.  The maximum image scale is 150~m/pix.
We divided the crater measurements into two regions of young, fractured terrain
(outlined in blue) and the heavily cratered terrain.}
\label{fig:encellimtop004}
\end{center}
\end{figure}

\clearpage

\begin{figure}
\begin{center}
\epsfxsize=100mm
\includegraphics[width=5truein]{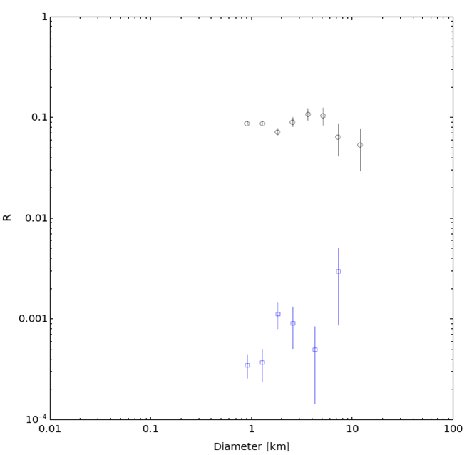}
\caption[Enceladus ISS\_003EN\_LIMTOP004\_PRIME.]
{The R-plot for the regions seen in Figure~\ref{fig:encellimtop004}.  The
measurements within the young, fractured terrain are the blue symbols,
while the measurements within the older, cratered terrain are the
black symbols.  We measured 48 craters in the fractured terrain,
and 1840 craters in the cratered terrain.
Again, the young terrain SFDs
here follow a $\sim -2$
differential slope.  The cratered terrain has a $\sim -3$
differential slope above a few km.  See text for discussion.}
\label{fig:encellimtop004rplot}
\end{center}
\end{figure}

\clearpage
\begin{figure}
\begin{center}
\epsfxsize=100mm
\includegraphics[width=5truein]{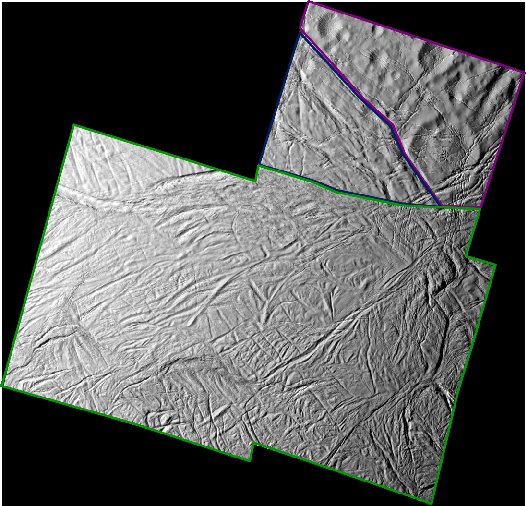}
\caption[Enceladus ISS\_011EN\_MORPH002\_PRIME.]
{The Enceladus
ISS\_011EN\_MORPH002\_PRIME mosaic.  The average
mosaic scale is 67~m/pix, and the maximum image scale is 80~m/pix.
We divided the crater
measurements into two regions of young, fractured terrain
(outlined in blue and green) and the heavily cratered terrain
(outlined in purple).}
\label{fig:encelmorh002}
\end{center}
\end{figure}

\clearpage

\begin{figure}
\begin{center}
\epsfxsize=100mm
\includegraphics[width=5truein]{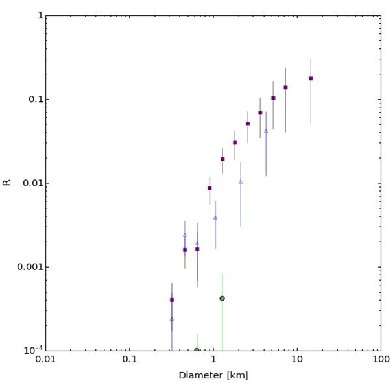}
\caption[Enceladus ISS\_011EN\_MORPH002\_PRIME.]
{The R-plot for the regions seen in Figure~\ref{fig:encelmorh002}.  The
measurements within the young, fractured terrain are the blue and green 
symbols,
while the measurements within the older, cratered terrain are the
purple symbols.  We measured 4 craters in the fractured region outlined
in green, 15 craters measured in the fractured area outlined in blue,
and 53 craters in the cratered terrain outlined in purple.
The young terrain SFDs here follow a $\sim -2$
differential slope.  The cratered terrain area is too small to
contain large craters with a spread of diameters, and so does not
have a distinct slope for diameters above several km.  However,
like the heavily cratered regions seen in previous figures, the
crater density decreases at smaller diameters.}
\label{fig:encelmorh002rplot}
\end{center}
\end{figure}

\clearpage

\begin{figure}
\begin{center}
\epsfxsize=100mm
\includegraphics[width=5truein]{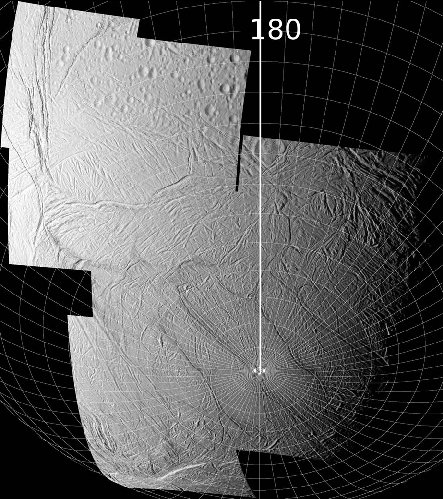}
\caption[Enceladus ISS\_011EN\_N9COL001\_PRIME.]
{The Enceladus
ISS\_011EN\_N9COL001\_PRIME mosaic.  The average
mosaic scale is 110~m/pix, and the maximum image scale is 129~m/pix.
We show a latitude/longitude grid on this mosaic to emphasize the
extent of the young terrain imaged in this mosaic, and the excellent
coverage of the south pole itself.  The 180 label refers to the
longitude of that line.  For this mosaic, we measured
only the young terrain.}
\label{fig:enceln9col001}
\end{center}
\end{figure}

\clearpage

\begin{figure}
\begin{center}
\epsfxsize=100mm
\includegraphics[width=5truein]{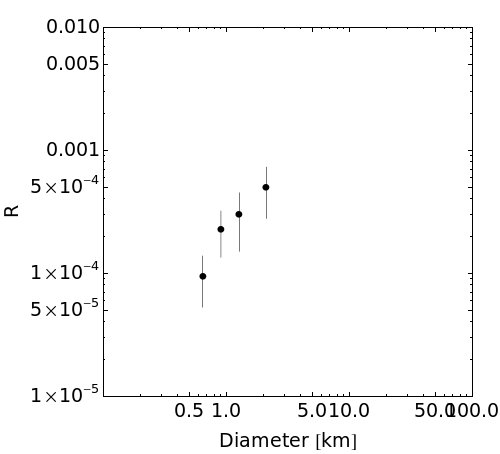}
\caption[Enceladus ISS\_011EN\_N9COL001\_PRIME.]
{The R-plot for the young terrain seen in Figure~\ref{fig:enceln9col001}.
We measured 22 craters in this region.  
The $\sim -2$ differential slope is similar with other measurements
within the fractured terrain, but at a lower density (lower R-value),
consistent with the expectation that the area around the south pole
is the youngest region on Enceladus.}
\label{fig:enceln9col001rplot}
\end{center}
\end{figure}

\clearpage

\begin{figure}
\begin{center}
\epsfxsize=100mm
\includegraphics[width=5truein]{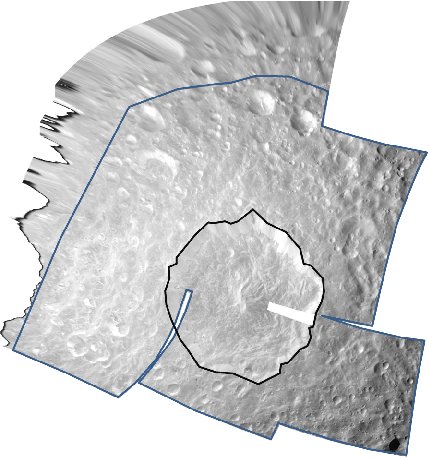}
\caption[Cassini ISS\_126MI\_GEOLOG001\_PRIME.]
{The Cassini ISS\_126MI\_GEOLOG001\_PRIME mosaic of Mimas, with an average
image scale of about 105~m/pix.  The large crater in this mosaic is
Herschel crater, about 135~km diameter.  The measurements we report are
for the area outside Herschel crater (in between the blue and black
outlines).}
\label{fig:mimasimage}
\end{center}
\end{figure}

\clearpage

\begin{figure}
\begin{center}
\epsfxsize=100mm
\includegraphics[width=5truein]{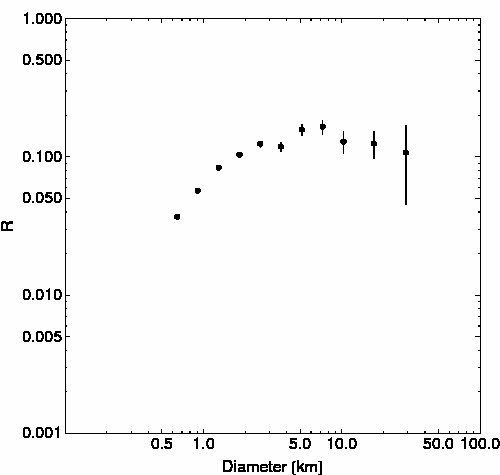}
\caption[Mimas r-plot.]
{An R-plot for our measurements of the area shown in
Figure~\ref{fig:mimasimage}; we measured 6581 craters in that region.  
The density decrease at crater diameters
less than about 10~km is not a completeness effect, because the average
mosaic scale is about 105~m/pix.  The decrease in small crater density
supports our calculation that secondary craters should be rare to
non-existent on Mimas. }
\label{fig:mimasrplot}
\end{center}
\end{figure}

\clearpage

\begin{figure}
\begin{center}
\epsfxsize=100mm
\includegraphics[width=5truein]{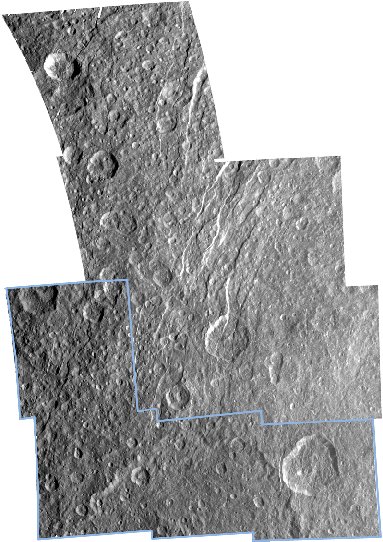}
\caption[Rhea mosaic.]
{A subset of the Cassini ISS\_121RH\_REGMAP001\_PRIME Rhea mosaic, average
mosaic scale of about 159~m/pix. The presence of a higher-density small
crater population supports our prediction that Rhea is sufficiently
massive to support the production of a secondary crater population.}
\label{fig:rheaimage}
\end{center}
\end{figure}

\clearpage

\begin{figure}
\begin{center}
\epsfxsize=100mm
\includegraphics[width=5truein]{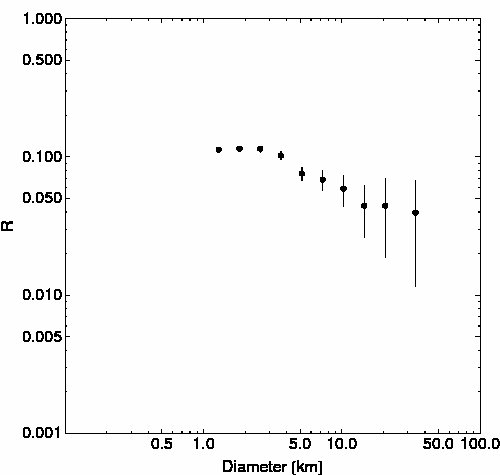}
\caption[Rhea r-plot.]
{An R-plot for our measurements of the area shown in
Figure~\ref{fig:rheaimage}; we measured 7495 craters in that
region.  Unlike Enceladus or Mimas, there is not a
crater density decrease at smaller diameters.}
\label{fig:rhearplot}
\label{lastfig}
\end{center}
\end{figure}

\end{document}